\documentclass[%
 reprint,
 amsmath,amssymb,
prb,
floatfix,
]{revtex4-2}
\usepackage{empheq}
\usepackage{float}
\usepackage{lipsum}
\usepackage{mathtools, cuted}
\usepackage{esvect}
\usepackage{xcolor}
\usepackage[caption = false]{subfig}
\usepackage{graphicx}% Include figure files
\usepackage{dcolumn}% Align table columns on decimal point
\usepackage{bm}% bold math
\usepackage{hyperref}% add hypertext capabilities
\usepackage{physics}
\usepackage[mathlines]{lineno}% Enable numbering of text and display math
\begin{document}

\preprint{APS/123-QED}

\title{Effects of Disorder in the Fibonacci Quasicrystal}% Force line breaks with \\

\author{Anouar Moustaj}
 \affiliation{Institute of Theoretical Physics, Utrecht University.}%
\author{Sander Kempkes}
\affiliation{Institute of Theoretical Physics, Utrecht University.}%
\author{Cristiane Morais Smith}
\affiliation{Institute of Theoretical Physics, Utrecht University.}%

\date{\today}% It is always \today, today,

\begin{abstract}
We study the properties of the one-dimensional Fibonacci chain, subjected to the placement of on-site impurities. The resulting disruption of quasiperiodicity can be classified in terms of the renormalization path of the site at which the impurity is placed, which greatly reduces the possible amount of disordered behavior that impurities can induce. Moreover, it is found  that, to some extent, the addition of multiple, weak impurities can be treated by superposing the individual contributions together and ignoring nonlinear effects. This means that a transition regime between quasiperiodic order and disorder exists, in which some parts of the system still exhibit quasiperiodicity, while other parts start to be characterized by different localisation behaviours of the wavefunctions. This is manifested through a symmetry in the wavefunction amplitude map, expressed in terms of conumbers, and through the inverse participation ratio. For the latter, we find that its average of states can also be grouped in terms of the renormalization path of the site at which the impurity has been placed. 

% This is then found to depend on the nature of the gap in which the impurity mode has most influence. Since these gaps are labelled by an integer Chern number through the normalized integrated density of states and the gap labelling theorem, we investigate the effect of breaking palindromic symmetry on the topological character of the system. 
% \begin{description}
% Secondary publications and information retrieval purposes.
% You may use the \texttt{description} environment to structure your abstract;
% use the optional argument of the \verb+\item+ command to give the category of each item. 
% \end{description}
\end{abstract}

\maketitle

%\tableofcontents

\section{\label{sec:Intro}Introduction %:\protect\\ The line
%break was forced \lowercase{via} \textbackslash\textbackslash
}

Since their discovery by Dan Shechtman \cite{shechtman}, quasicrystals have attracted much attention. Their unusual properties, such as low thermal conductivity, low friction coefficients, high hardness, corrosion resistance and superplasticity have made them attractive for applications. Their utility ranges from heat-insulating materials, through coating that increases hardness, all the way to medical implants, where prosthetics made from quasicrystalline material have shown very little cytotoxicity effects \cite{popeconduc,alpdreelec,quasifeaturesbook,emaciaaporder}. Many studies have been conducted on the nature and properties of quasicrystals. Not being identified as classical crystals, they are characterized by forbidden discrete symmetries, such as the fivefold rotation group, with the archetypal example being AlMg \cite{shechtman}. One way to understand the emergence of this symmetry is by tiling a 2D plane \cite{DEBRUIJN198153}. On the other hand, a more structural way of understanding all quasiperiodic arrangements is to view them as projections from higher dimensional lattice spaces, which have a perfectly periodic structure \cite{tedjanssen,Janssen:a25379}. Although mainly manufactured in laboratories, quasicrystals have also been observed to occur in nature, where the structure was found to exist in a Siberian meteorite sample \cite{Bindi1306}. A mechanism for the formation of quasicrystalline phases, both artificial and in nature, has recently been proposed. It consists of the superposition of two 1D periodic subsystems with incomensurate periods, where charge-density waves favor the emergence of a quasiperiodic tiling of the atomic lattice \cite{CDWQUASIPER}.

The fact that quasicrystals are not periodic in their microscopic structure makes them a more complicated problem to study than their periodic counterpart. One way to simplify the problem and still obtain relevant results is to study a 1D abstraction of the real system. The most popular toy model in that case is given by the 1D Fibonacci quasicrystal \cite{NORI}. This is a tight-binding model for a particle subject to a lattice potential. Either the on-site potential or the hopping parameter, depending on the model chosen, is modulated by the Fibonacci sequence and takes on two discrete values, as will be explained in more detail later on. This model exhibits very interesting properties: its energy spectrum is singular continuous, which makes its semi-infinite version a proper fractal set, and its wavefunctions possess multifractal properties \cite{KOHO}. A renormalization scheme was introduced to explain the features of the spectrum and its scaling symmetries \cite{NORI}. This scheme was subsequently used to understand the gap labeling theorem \cite{Bellissard1992}, applied to the purely hopping Fibonacci chain. Together with the conumbering scheme \cite{SireMosseri}, it offered an insightful way to characterize the wavefunctions in terms of their renormalization paths \cite{Mac__2016}. Nowadays, new insights are still being provided. These range from the topological character of the system to superconductivity \cite{levy2016topological,Mac__2016,R_ntgen_2019,Kraus_2012,krausexp,PhysRevB.100.165121,SupCondOPfluc}. 

In this paper, we aim at understanding how impurities disrupt the quasiperiodic order. We start by investigating the effect of a single impurity on the wavefunction canvas using the aforementioned renormalization scheme. We find that for weak impurities, a transition regime exists in which the quasiperiodic order remains intact in parts of the system. This is manifested by the preservation of a symmetry in the wavefunction amplitude as a function of conumbered sites and can be quantified by calculating cluster-averaged overlap integrals. As the strength of the impurity is raised, a disordered phase appears, which is characterized by the localization of the wavefunctions. This is compatible with previous results in Ref.~\cite{subsdisorder}, where ``resonant states'' were identified in the presence of an impurity. Moreover, we find that the transition regime can be labelled by the renormalization path of the site at which the impurity has been placed. This is also visible in the change of the state-averaged inverse participation ratio (IPR) as a function of the impurity strength. The different behaviors of the IPR can be grouped in terms of the renormalization path of the site at which the impurity has been placed. Some impurity realizations lead to the surpising delocalization-relocalization phenomena for a range of impurity strengths. This observation is very similar to one found in a recent study, where random disorder was introduced in the hopping parameters and was shown to lead to a regime of delocalization before a ``reentrant localization'' takes place \cite{Anuradha,NonmonotonicXover}. To some extent, this also holds when adding multiple weak impurities, where the individual contributions can just be superposed to produce the full disruptive pattern. Since the quasiperiodic order is gradually lost, the transition regime can be characterized by a classification of the kind of disorder induced, and we find that disorder develops in an organized way. 

The paper is structured as follows. In Sec.~II, we give a brief description of the Fibonacci chain in a tight-binding approximation. We then provide an overview of the understanding of its spectrum through a renormalization procedure. In Sec.~III, we introduce disorder by adding one impurity to the system and show that we can classify the kind of disorder by the renormalization path label of the site at which the impurity has been placed. Finally, in Sec.~IV, we present our conclusions and outlook.

\section{\label{sec:Themodel} The Model}
We start by briefly introducing the Fibonacci chain in the tight-binding approximation, followed by an analysis of its properties through a renormalization procedure. 
\subsection{\label{sec:FibTBM}Fibonacci Tight-Binding Model}
\subsubsection{Fibonacci Sequence}

\paragraph{Inflation Method}
The Fibonacci sequence, represented by a binary alphabet $\{L,S\}$, can be generated iteratively through the inflation rule 
\begin{equation*}
    \begin{split}
        S\to L, \\
        L \to LS,
    \end{split}
\end{equation*}
starting with the ``zeroth'' letter $S$. The $N^{\text{th}}$ iteration, $W_N$, will be referred to as the $N^{\text{th}}$ approximant of the Fibonacci word, the size of which will be denoted by $|W_N|=F_N$. This sequence has the property 
\begin{equation}
    \lim_{N\to\infty}\frac{F_{N+1}}{F_N}=\frac{1+\sqrt{5}}{2}\equiv \phi,
\end{equation}
where $\phi$ is called the \textit{golden ratio}. The Fibonacci sequence is often represented in terms of word-size and takes the form $\{F_N\}_{N=0}^\infty=\{1,1,2,3,5,8,13,\dots\}$. Another property is that each term can be generated recursively through: 
\begin{equation}
    F_{N+2}=F_{N+1}+F_N,
\end{equation}
with $F_0=F_1=1$. In terms of Fibonacci words, the recursion relation can be written as
\begin{equation*}
    W_{N+2}=W_{N+1}W_N.
\end{equation*}
\begin{figure*}[!hbt]
    \centering
    \includegraphics[width=\textwidth]{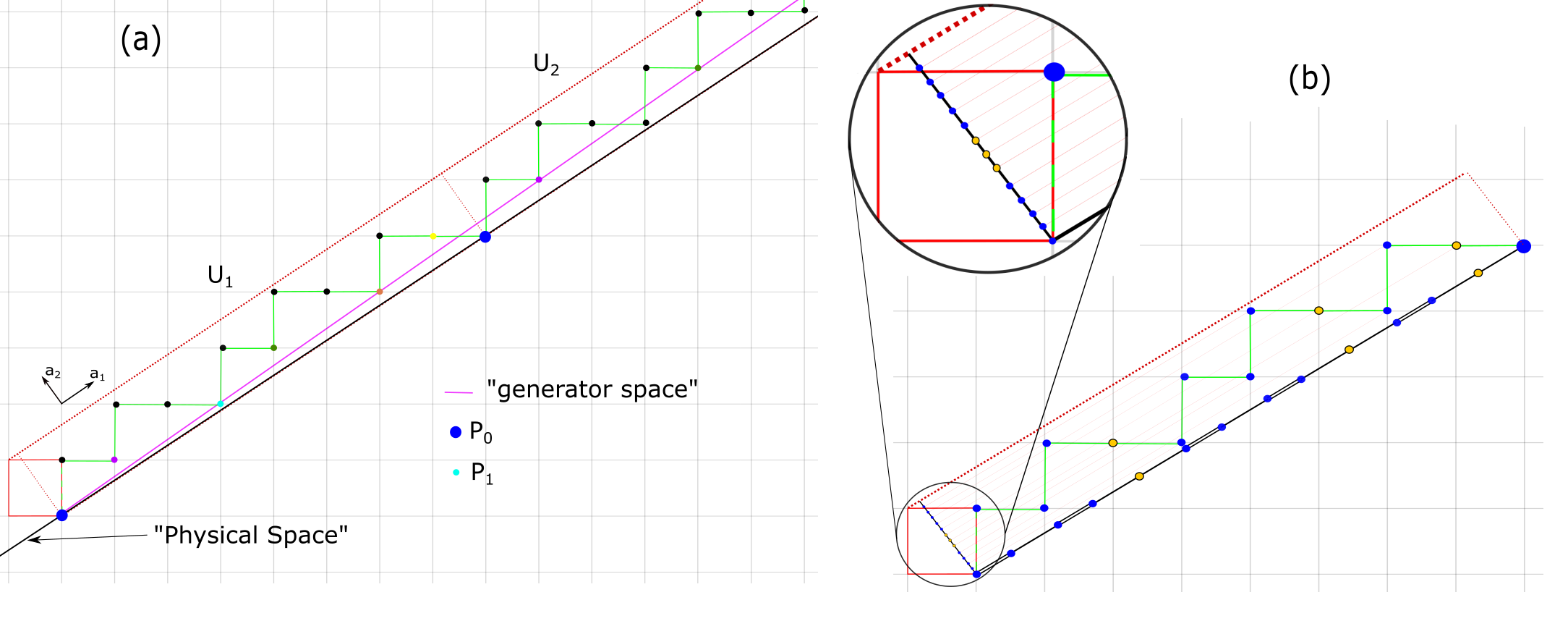}
    \caption{(a) $\mathbb{Z}^2$ space, with a Fibonacci quasicrystal
    approximant cell containing 13 sites ($\phi_6=8/5$). The vectors $\mathbf{a}_1$ and $\mathbf{a}_2$ are also show for reference. The strip containing the sites to be projected is defined by translating the red square along the physical space. The repeating unit cells are denoted by $U_i$. The purple line is the generator space span$\{\mathbf{h}\}$, with $\mathbf{h}=(3,2)$. The points on the generator are marked with different colors. Notice that points in unit cell $U_2$ below the generator line are just repetitions of the points already marked in the first unit cell $U_1$ (in this case, the cyan and orange points). (b) Both projections are shown here, with a zoom in on the orthogonal space to show how the conumbered sites are organized. The real space chain has also been adorned with the different hoppings corresponding to the two distances (L and S).}
    \label{cutproj}
\end{figure*}
\paragraph{Cut and Project Methods and Conumbers}
The Fibonacci sequence is known to be quasiperiodic, i.e. it can be obtained as a projection of a higher-dimensional periodic sequence. This is the so called cut-and-project method of generating quasiperiodic lattices. In this case, points on $\mathbb{Z}^2$ are projected onto a line of slope $1/\phi$ (see Fig.~\ref{cutproj} (a), where the line represented in black is called the ``physical space''). In order to construct the $N^\text{th}$ periodic (rational) approximant of the infinite chain, we define the vector $\mathbf{a}_1=(F_{N-1},F_{N-2})$, which points along the line of slope $1/\phi_N=F_{N-2}/F_{N-1}$ (with $\phi_{N\to\infty}=\phi$). We then consider a unit square, the lower left vertex of which lies at the origin of the chain [see the red square in Fig.~\ref{cutproj} (a)], which is then translated in the direction of $\mathbf{a}_1$. All the points within the strip of width $\sqrt{2}$ traced by the square are then projected onto the physical line. The result is a chain in which nearest neighbours have relative distances $L$ (long) and $S$ (short), arranged in a Fibonacci sequence. This construction is enough to obtain the Fibonacci chain itself. However, we will also consider the projection in the orthogonal space, spanned by the vector $\mathbf{a}_2=(-F_{N-2},F_{N-1})$. When the same selection of points within the strip is projected along the line spanned by $\mathbf{a}_2$, a very interesting arrangement of the sites is obtained. Indeed in Fig.~\ref{cutproj}(a) we see that the projection along $\mathbf{a}_2$ amounts to an ordering of the sites in terms of their distance from the physical line. The shortest distance corresponds to the cyan point, followed by the orange point, the magenta point and so on. We observe that they all lie on the line called ``generator space'' (the origin of the name will become clear later). The Fibonacci chain allows for two types of sites: the ones surrounded by two $L$ bonds, which we shall call ``atomic'' and represent by a yellow dot, and those separated by an $S$ bond, henceforth called ``molecular'' and represented by a blue dot (see Fig.~\ref{cutproj} (a) and (b)). In Fig.~\ref{cutproj}(b), we see that on top of rearranging the points according to their distance from the physical line, the orthogonal projection also rearranges them in terms of their type: the atomic sites being placed in the middle, while the molecular ones are at the sides of the newly arranged chain. This scheme is called ``conumbering'', as the rearrangement of the sites is achieved by attaching a ``conumber'' to each site of the original Fibonacci chain. The scheme is constructed as follows: we define a ``generator'', $\mathbf{h}=\vv{P_0P_1}$, where $P_0$ and $P_1$ are the origin and the point with the smallest distance to the physical line. As stated previously, all subsequent points of shortest distances will lie on the line generated by $\mathbf{h}$. However, these points will not all lie within the first unit cell [denoted $U_1$ in Fig.~\ref{cutproj}(a)], and the main function of the conumbering scheme is to identify the points that lie outside of the first unit cell with the points that lie in it. This is achieved by performing a modulo operation on those points. The generator $\mathbf{h}$ will always generate the nearest previous rational approximant line that lies within the strip traced by the unit square. This construction can be visualized in Fig.~\ref{cutproj}(a). Thus conumbered points, before projection are given by 
\begin{align*}
    \mathbf{x}_{j}&=j\mathbf{h}\text{mod}[\mathbf{a_1}] \\
    &=j\bigg(a\text{mod}(F_{N-1}),b\text{mod}(F_{N-2})\bigg)
\end{align*}
where $a$ and $b$ are two Fibonacci numbers that identify the nearest point and $j\in\{0,\hdots,F_{N}-1\}$ is the ``conumber''. There is a periodicity in the difference between consecutive rational approximants of $\phi$: 
\begin{equation*}
    \text{sign}(\phi_{N+1}-\phi_N)=(-1)^{N+1}, \ \ N=\{0,1,2,...\}.
\end{equation*}
This means that the slope that determines the generator is either the previous approximant or the one before it, depending on whether this slope is higher or lower than the actual approximant.  After projection, this yields the usual formula used in recent literature \cite{SireMosseri,Mac__2016},
\begin{equation}\label{conumbers}
    j=x_jF_{N-1}\text{mod}(F_N),
\end{equation}
where $x_j\in\{0,\hdots,F_{N}-1\}$ denotes the sites in increasing order on the real lattice. The conumbered sites are very convenient because they organize the real lattice sites in terms of their local environment. The other advantage is that one can observe the symmetry between the energy levels and the amplitude localization (see Sec.~III).

\subsubsection{Tight-Binding Hamiltonians}
The Fibonacci chain is constructed by considering the nearest-neighbor tight-binding Hamiltonian 
\begin{equation}\label{Fibohamilt}
    H=\sum_{i=1}^{\infty}\bigg[V_i\ket{i}\bra{i}+t_i\ket{i}\bra{i+1}+ \text{h.c.}\bigg].
\end{equation}
We can either modulate the on-site potential $V_i$ or the hopping parameter $t_i$. We shall refer to the two cases as the ``on-site model'' and the ``hopping model'', respectively. The modulation is applied as follows: 
\begin{equation}
    V_i=\begin{cases}
               V_w, \ \ \text{if $i^\text{th}$ letter is $L$}, \\
               V_s, \ \ \text{if $i^\text{th}$ letter is $S$},
            \end{cases}
\end{equation}
for the on-site model and 
\begin{equation}
    t_i=\begin{cases}
               t_w, \ \ \text{if $i^\text{th}$ letter is $L$}, \\
               t_s, \ \ \text{if $i^\text{th}$ letter is $S$},
            \end{cases}    
\end{equation}
for the hopping model. The subscripts $w$ and $s$ have been chosen to reflect that $L$, standing for long, would correspond to a weak bond (hopping strength) and $S$ (short) for a stronger bond. The two models have been studied extensively in Refs.~\cite{NORI,R_ntgen_2019,Pichon}, where the renormalization scheme \cite{NORI} was used to reveal the multifractal properties of the model \cite{Mac__2016,Pichon}. The on-site model was also studied in Ref.~\cite{R_ntgen_2019} through the perspective of local symmetries, where a systematic way to control the edge modes of a finite chain was devised.

\begin{figure}[!hbt]
    \includegraphics[width=0.9\columnwidth]{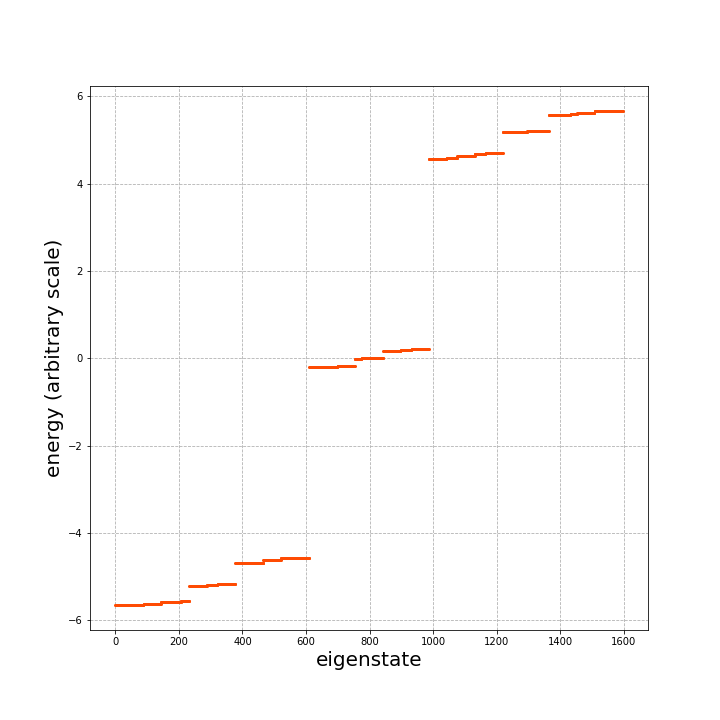}
    \caption{\label{fig:1597} Energy spectrum of the N=16 approximant Fibonacci chain, with 1597 sites, in the hopping model. The trifurcating structure can be seen at different energy scales. The self-similarity of this structure is also visible. The modulation strength has been set to $\rho\equiv t_w/t_s=0.2$.}
\end{figure}

In the remainder of this paper, we focus on the hopping model, to which we apply periodic boundary conditions to properly renormalize it. The resulting spectrum has many interesting properties, which are characteristic of quasiperiodic systems. The semi-infinite chain is singular continuous and is also a fractal \cite{Mac__2016,fractalmeasures}. An example of the spectrum of a $N=16$ chain is shown in Fig.~\ref{fig:1597}. There, we observe a trifurcarting structure that is self-similar. A renormalization procedure was devised by Nori et al.~\cite{NORI} to explain these features. There is also a direct mapping between the on-site and hopping models under the perturbative renormalization scheme (see Ref.~\cite{NORI}).  
\subsection{\label{sec:RNMandLRM}Renormalization of the Chains}

The spectrum of the Fibonacci chain can be understood by performing a perturbative renormalization procedure to the hopping model, which is exact in the limit $\rho\equiv t_w/t_s\to0$. Note that this can also be applied to the on-site Fibonacci chain, which after one renormalization step becomes a hopping Fibonacci chain. We start from the original Hamiltonian and without loss of generality, set the on-site energy to be a constant, $V_i\equiv V=0$. Then, we split it into an unperturbed part $H_0$ and a perturbation $H_1$, with 
\begin{equation}
    \begin{split}
        H_0&=\sum_j t_s\ket{j}\bra{j+1} +\text{h.c,\ \ \ if }\ \ j\ \text{mod}(\phi)<\phi^{-1}, \\
        H_1&=\sum_j t_w\ket{j}\bra{j+1} +\text{h.c,\ \ \ if }\ \ j\ \text{mod}(\phi)\geq\phi^{-1}.
    \end{split}
\end{equation}
The conditions imposed rely on the quasiperiodicity of $\text{sign}\big[(j+1)\text{mod}(\phi)-j\text{mod}(\phi)\big]$, which corresponds to the Fibonacci sequence in terms of $\{-1,1\}$.

\begin{figure}[!hbt]
    \includegraphics[width=0.9\columnwidth]{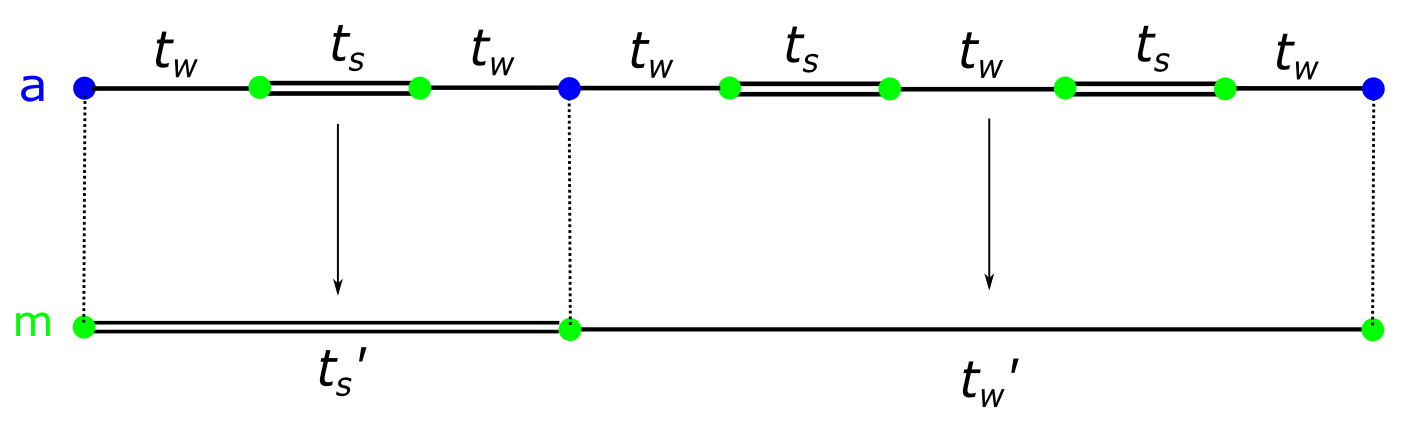}
    \caption{\label{fig:atomicdec}(Color online) Decimation of the atomic-leveled chain. We have also colored the sites consistently with our previous definition for atomic and molecular sites. This figure gives a simple example of a renormalization path of a site in the $N^\text{th}$ approximant chain. In this case, a chain of length $F_5=8$ is renormalized to a chain of length $F_2=2$. The renormalization path is just `am'. This figure was inspired by those made in Refs.~\cite{NORI,Mac__2016}.}
\end{figure}

\begin{figure}[!hbt]
    \includegraphics[width=0.9\columnwidth]{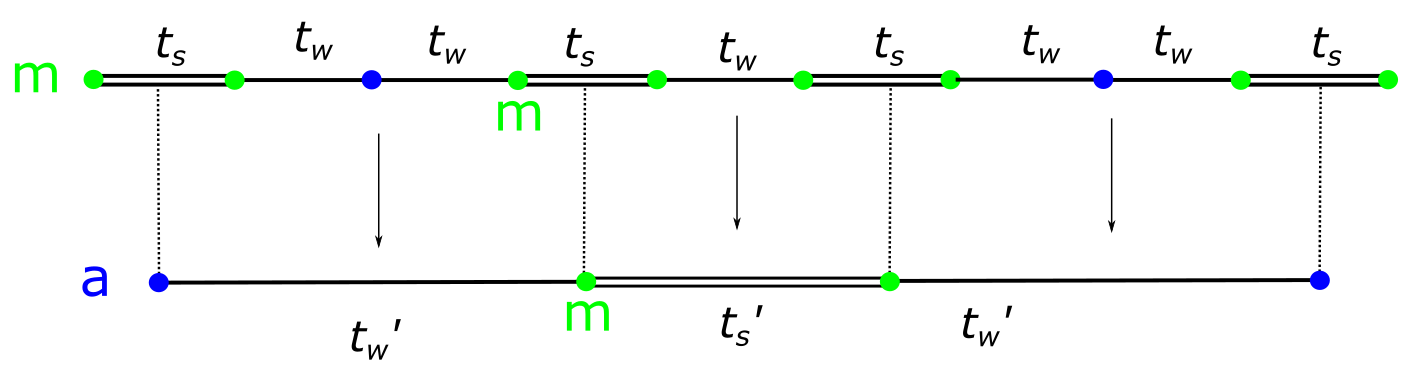}
    \caption{\label{fig:molecdec}(Color online) Decimation of the molecular-leveled chain. The same coloring scheme has been used to label the sites. Here, we have another simple example of renormalization paths, namely `ma' on the left and `mm' in the middle. The chain of length $F_5=8$ is changed to a chain of length of $F_3=3$. This figure was inspired by those made in Refs.~\cite{NORI,Mac__2016}.}
\end{figure}

The unperturbed Hamiltonian has three levels with a very large degeneracy, namely $E=0,\pm t_s$. This sets the starting point of the renormalization procedure, which is applied to each of the three unperturbed levels independently. Following the nomenclature proposed by Macé et al.~\cite{Mac__2016}, we have one atomic deflation, corresponding to the atomic level ($E=0$) and two molecular deflations corresponding to the bonding and anti-bonding molecular levels ($E=\pm t_s)$. The atomic deflation takes the original chain of size $F_N$ to a smaller Fibonacci chain of size $F_{N-3}$, while the molecular deflations map the original chain to one of size $F_{N-2}$. The renormalized hopping strengths, in each case, are given by:
\begin{empheq}[left = {\{t_w',t_s'\}= \empheqlbrace}]{align*}
               & \left\{\frac{t_w^2}{2t_s},\frac{t_w}{2}\right\}=\frac{\rho}{2}\{t_w,t_s\}, \ \ \ \text{(bonding)} \\
               & \left\{\frac{t_w^3}{t_s^2},-\frac{t_w^2}{t_s}\right\}=\rho^2\{t_w,-t_s\}, \ \ \ \text{(atomic)} \\
               & \left\{\frac{t_w^2}{2t_s},-\frac{t_w}{2}\right\}=\frac{\rho}{2}\{t_w,-t_s\}, \ \text{(anti-bonding)}.
\end{empheq}
We can thus write the original $N^\text{th}$ approximant Hamiltonian, $H_N$, as a direct sum of three sub-Hamiltonians up to and including $\mathcal{O}(\rho^3)$ \cite{Pichon}: 
\begin{equation}
    H_N=\left(\frac{\rho}{2}H_{N-2}+t_s\right)\oplus\rho^2H_{N-3}\oplus\left(\frac{\rho}{2}H_{N-2}-t_s\right).
\end{equation}
Since each of the chains are themselves Fibonacci approximants, we can apply this procedure iteratively until one cannot decimate any generation further. The procedure for each type of cluster is depicted in Figs.~\ref{fig:atomicdec} and \ref{fig:molecdec}.
To be more precise, the atomic decimation takes every atomic site of the chain ($F_{N-3}$ of them) and connects every pair with a renormalized hopping strength that is strong when they are close to each other, and weak when they are separated by longer distances (see Fig.~\ref{fig:atomicdec}). The molecular decimation, on the other hand, takes every superposition of two molecular sites (eigenstate of energy $\pm t_s$, each of which has a $F_{N-2}$ degeneracy), connected by $t_s$ in the original Fibonacci chain, and creates a new chain made up by these superpositions. The hoppings are connected in a similar way as in the atomic deflation, i.e. a strong hopping whenever these two renormalized sites are separated by a short distance in the original chain, and a weak hopping when the distance is larger (see Fig.~\ref{fig:molecdec}). 

The notion of renormalization path will play an important role later on, when we study the effect of impurities in a Fibonacci chain. The renormalization path is defined in two ways. The first one is a string of letters `a' and `m', which stand for the nature of the site at each renormalization step. Starting from an $N^\text{th}$ approximant chain, two types of paths appear: `amma...’ for an atomic site at the top of the renormalization process, or ‘mmam...’ for a molecular one. The other definition pertains to energy eigenstates and the cluster to which they belong. Since there are three distinct clusters, we define the eigenstate renormalization path by a string of letters ‘t’, ‘c’ and ‘b’, describing the top cluster (bonding molecular), the central cluster (atomic) and the bottom cluster (anti-bonding molecular), respectively. A particular level can then be encoded by the symbolic string sequence ‘tctbt...c’ for example.  There exists a particular symmetry between the two renormalization paths in the perturbative limit ($\rho\ll1$), which has been made evident by the graphical representation given in Ref.~\cite{Mac__2016}. This symmetry manifests itself when we rearrange the chain sites in terms of the conumber defined in Sec.~II $C_N(j)=jF_{N-1}\text{mod}(F_N)$ [see Fig.~\ref{cutproj}(b)]. The fractal nature of the eigenstates becomes manifest in this representation, which is used in Sec.~III to show how impurities disrupt this order.
% \begin{figure}
%     \centering
%     \includegraphics[width=\columnwidth]{N=55_classnoimp.png}
%     \caption{Wavefunction amplitude map as a function of conumber (color online). On the $x$-axis, we show the conumbers and for reference, we also show the real lattice sites. They are colored by the nature of the site (green for molecular and blue for atomic), as done before in Figs.~\ref{fig:PERPCUT}, \ref{fig:atomicdec} and \ref{fig:molecdec}. We note the symmetry between the conumbers and the energy eigenstates, which are ordered in terms of increasing energy eigenvalues. In the quasiperiodic limit $N\to\infty$. This graph becomes a proper fractal.}
%     \label{fig:noimp}
% \end{figure}

\section{\label{sec:disorder}Introducing Disorder}

There are multiple ways of introducing disorder by adding impurities. It was already shown \cite{subsdisorder} that even the introduction of one impurity in a relatively long chain has a drastic effect on the spectrum. In particular, it reduces the fractal dimension of the global density of states, while increasing that of the local wavefunctions. Moreover, since the spectrum of the Fibonacci chain is singular-continuous \cite{fractalmeasures}, i.e. it has an infinite number of gaps, every state is affected by the presence of this impurity. The extent to which they are affected will however depend on the strength of the impurity.
We will see that in fact, for weak impurity strengths, departure from criticality can be characterized using the renormalization path formulation desribed before. To this end, we first show that we can organize the impurity-induced disorder by using the conumbering scheme. We then quantify this more thoroughly by studying the localization properties of the states through their inverse participation ratios. For the rest of the paper, we consider impurities by adding an extra term to the Hamiltonian: 
\begin{equation}
    H=H_F+\sum_mV_I\ket{m}\bra{m},
\end{equation}
where $H_F$ is the pure Fibonacci Hamiltonian and the impurities lie on a set of sites $\{m\}$, with $V_I$ denoting the impurity strength. We mainly focus on the single impurity case, but extend the discussion later on to multiple impurities of the same strength. 

\begin{figure*}[!hbt]
    \centering
    \includegraphics[width=.85\textwidth]{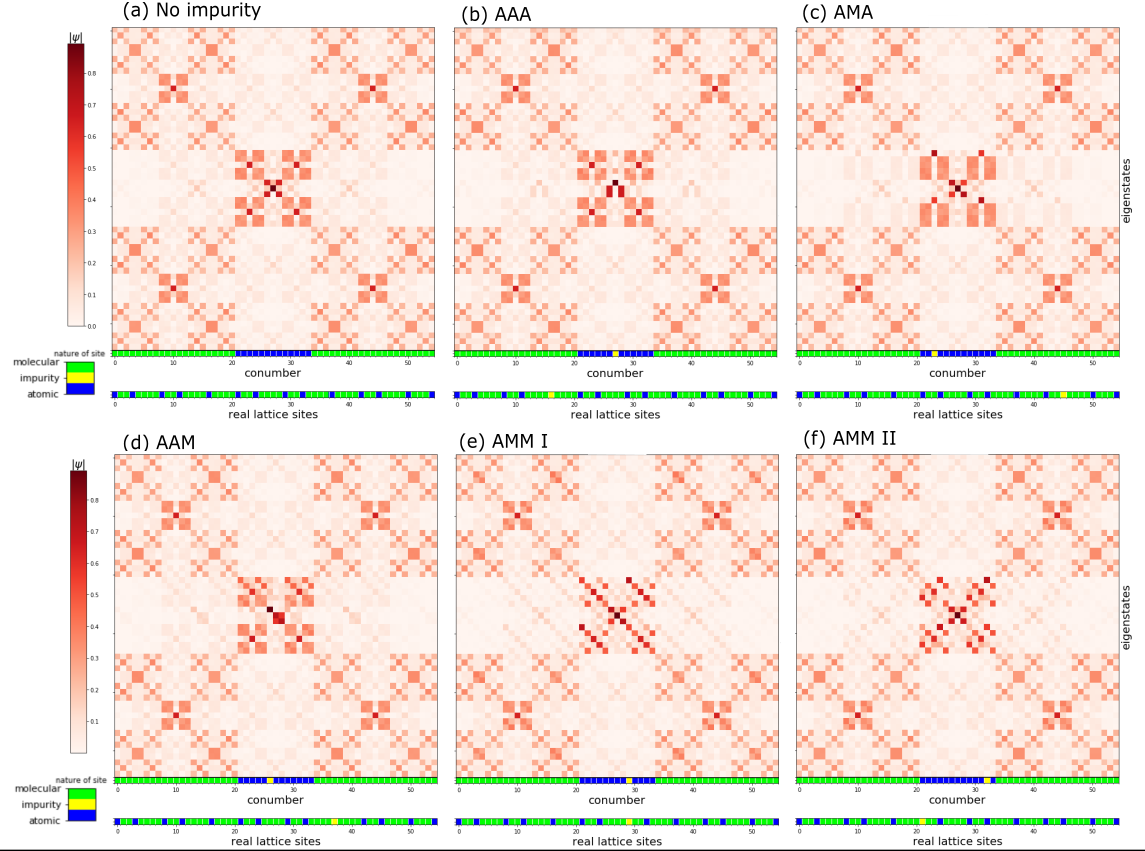}
    \caption{(Color online) Atomic disorder classes listed in table \ref{tab:RPDISORDER}. On the left side, we have two color codes: first, the local wavefunction modulus is plotted in shades of red; on the lower part, we have represented the lattice sites by their nature. Each site is either molecular (green), atomic (blue), or has an impurity (yellow). The parameters chosen for the numerical calculation are $\rho=0.2$ and $V_I=0.1t_w$. Note that because the impurity is weak, it mainly affects the atomic cluster and leaves the molecular cluster almost intact. (a) No impurity; (b-d) impurity on an AAA, AMA and AAM site, respectively;  (e) and (f) impurity on an AMM site, }
    \label{fig:classesATO}
\end{figure*}
\subsection{Organizing Disorder: Renormalization Path}

\begin{figure*}[!hbt]
    \centering
    \includegraphics[width=.85\textwidth]{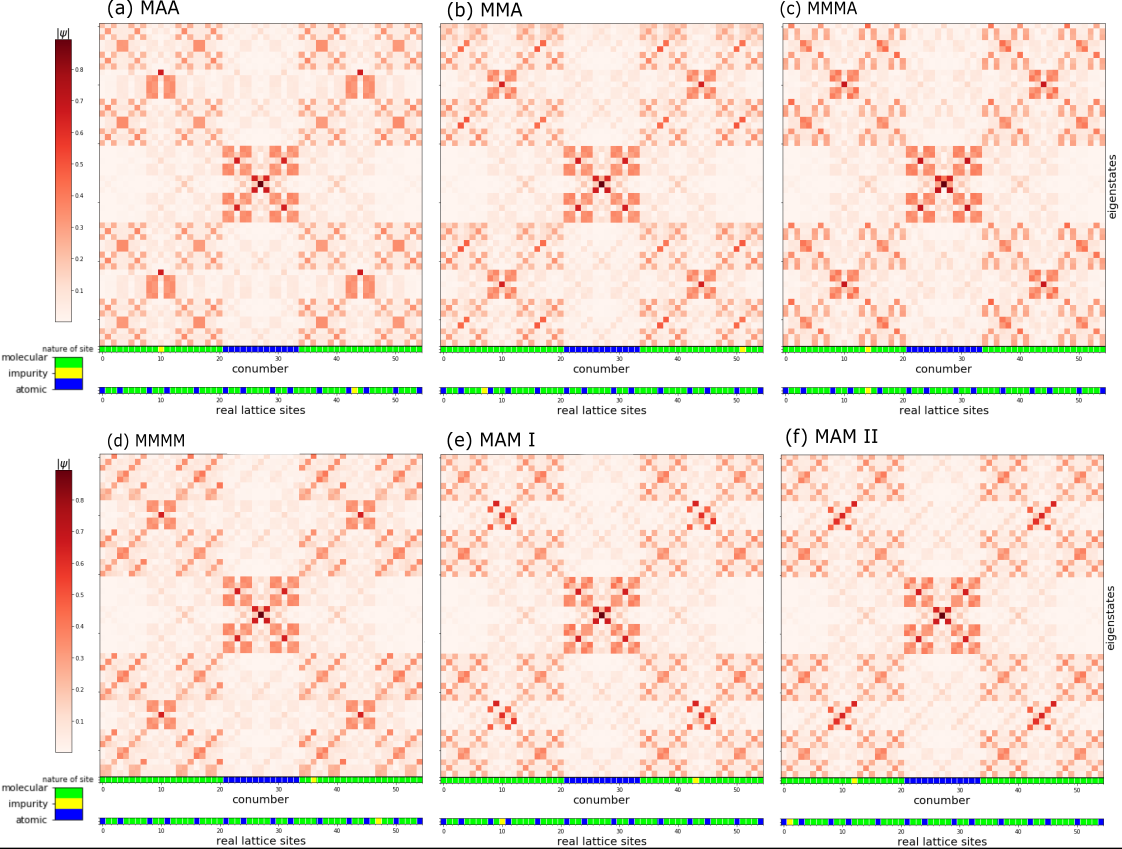}
    \caption{(Color online) Molecular disorder classes listed in table \ref{tab:RPDISORDER}. The color coding and parameters are the same as above. In this case, we find disorder mostly in the molecular cluster while the atomic one is left almost intact. (a-f) Impurities on sites belonging to classes MAA, MMA, MMMA, MMMM and MAM (I and II), respectively.}
    \label{fig:classesMOL}
\end{figure*}

Before one departs completely from criticality, it is possible to observe a structure in the way how disorder sets in. We found that it can be organized according to the renormalization path that the site, at which the impurity has been placed, follows. This is best exemplified through the concrete application of a weak impurity (at 10\% hopping strength) in a generation 9 chain, with 55 sites. In that case, a total of 9 renormalization paths exist. Each one of these is responsible for the generation of one disordered graph, as shown in Fig.~\ref{fig:classesATO} and Fig.~\ref{fig:classesMOL}.
From now on, we will positioned the impurities at
random sites in the Fibonacci chain. Then, we will identify the corresponding impurity positions in terms of conumbers. This allows us to identify which positions affect the eigenstate map the least. In Fig.~\ref{fig:classesATO}(a), we present the unperturbed fractal pattern, as obtained in Ref.~\cite{Mac__2016}. Its central part results mainly from the atomic sites, while the four surrounding structures result from molecular sites. Then, we successively position an impurity at various atomic sites in Figs.~\ref{fig:classesATO}(b-f) The resulting patterns are in an almost one-to-one correspondence with the renormalization paths of the sites at which the impurities are placed. Every site belonging to one path gives rise to the same disruption pattern [with the exception of the last two Figs.~\ref{fig:classesATO}(e),(f)]. The same is true in the case of impurities placed on molecular sites, as shown in Figs.~\ref{fig:classesMOL}(a-f). There is again a one-to-one correspondence between the pattern and the renormalization path, except for the last two cases, Figs.~\ref{fig:classesMOL}(e) and (f). 
\begin{table*}[!hbt]
\caption{\label{tab:RPDISORDER}List of renormalization paths and the amount of distinct graphs it generates. For reference we included the number of sites belonging to a particular renormalization path. Note the almost one-to-one correspondence between the number of graphs and the renormalization paths.}

\begin{ruledtabular}
\begin{tabular}{ccc}
 Renormalization Path&Number of sites&Number of distinct graphs\\ \hline
 MMMM& 16 & 1 \\
 MMMA& 8 
 & 1 \\
 MMA& 8& 1 \\
 MAM& 8 & 2 \\
 MAA& 2 & 1\\
 \hline
 AMM & 8 & 2\\
 AMA & 2 & 1\\ 
 AAM & 2 & 1\\
 AAA & 1 & 1
\end{tabular}
\end{ruledtabular}
\end{table*}
Starting from an atomic impurity at the carefully selected AAA site, as shown in Fig.~\ref{fig:classesATO}(b), we see that it mostly affects one eigenstate. Noting that the eigenstates are ordered in terms of increasing energy, we see that the slight increase from the impurity potential shifts this state's energy (which is the closest to the unperturbed atomic energy $E=0$) upwards. Since all of the renormalized chains correspond to atomic subclusters, the molecular clusters are left completely intact. In fact, even the molecular subclusters of the original atomic cluster are relatively well preserved. In Fig.~\ref{fig:classesATO}(c), the impurity is placed at a AMA site. We now have two states that are mostly affected by the presence of the impurity. These are the atomic states of the two $F_4$ chains resulting from the decimation procedure (one atomic site per chain, with atomic energies $E=\pm t_s'=\pm(\rho/2)t_s$, respectively). The previous two examples were the ones that disturbed the spatial distrubtion of the wavefunctions the least. The next three (AAM, AMM I and AMM II) show a higher level of disorder in the structure. However, we still see that in the weak impurity regime, this disorder is mainly confined to the atomic cluster. 

When placing an impurity on molecular sites, all resulting possiblities are shown in Fig.~\ref{fig:classesMOL}(a)-(f). In Fig.~\ref{fig:classesMOL}(a), we see the least amount of disorder, which can easily be explained, just as in the atomic case, by the disruption of two particular states. These are the atomic sites of two $F_4$ chains (one per chain, of energy $E=\pm t_s$, respectively), resulting from an MAA decimation procedure. This is qualitatively very similar to the AMA case, except that the atomic subclusters are different, and are centered around the unperturbed molecular energy states.  The next five graphs show the same kind of behavior, in the weak impurity regime, as in the atomic case. That is, we mostly see a disruption of the symmetric patterns within the molecular clusters, while the atomic cluster is mostly left intact.

We have listed in table \ref{tab:RPDISORDER} all the renormalization paths and number of graphs that they generate. The observation is that there is an almost one-to-one correspondence between the amount of renormalization paths and the possible type of disorder the system is subjected to. It is not exact, as there are two renormalization paths that each give rise to two disctinct graphs, namely AMM in the atomic case, and MAM in the molecular case [see Figs.~\ref{fig:classesATO} and \ref{fig:classesMOL} (e) and (f)]. Therefore, we can predict how many types of disordered graphs can be obtained if we can calculate the amount of distinct renormalization paths. 
\begin{figure}[!hbt]
    \centering
    \includegraphics[width=\columnwidth]{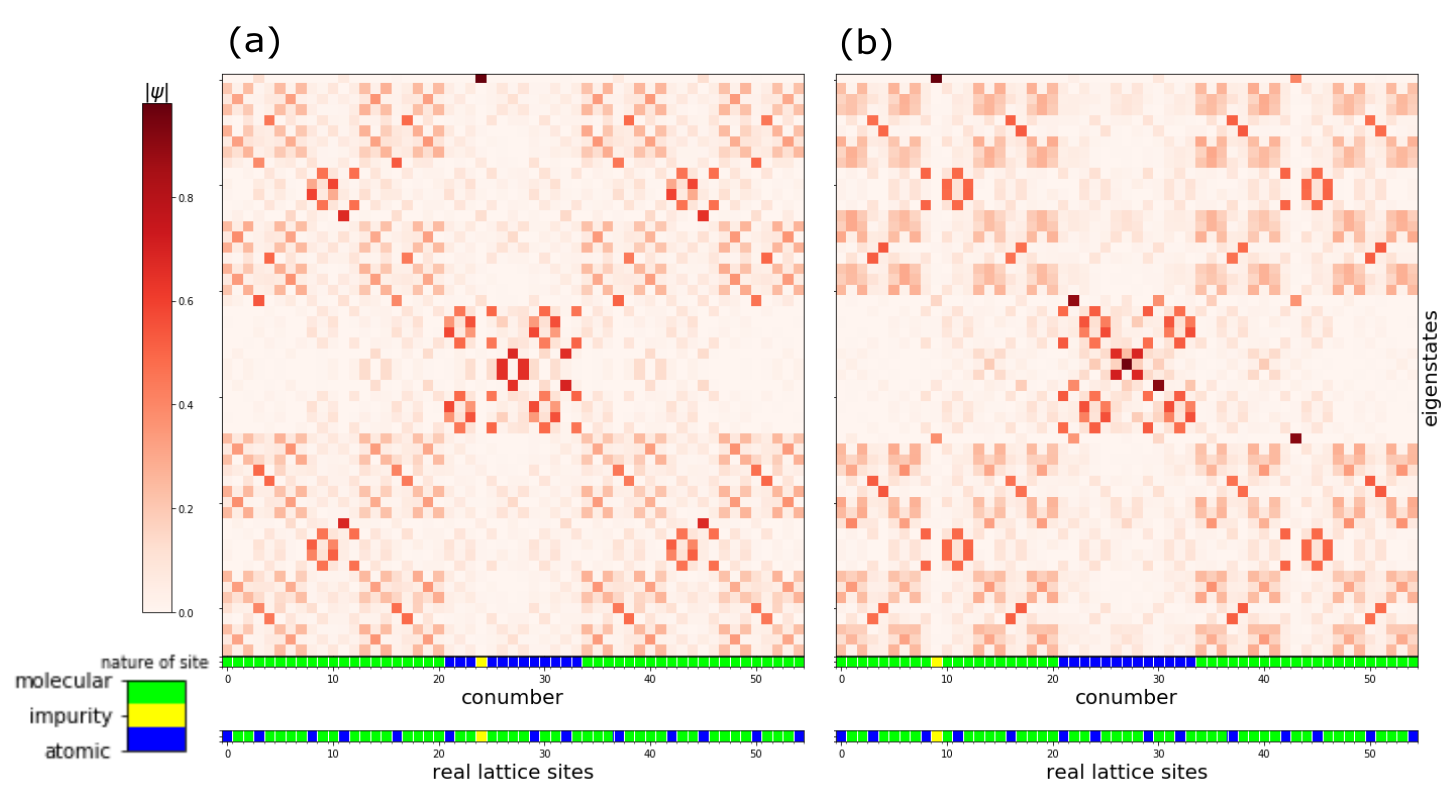}
    \caption{(Color online) Disorder induced by a strong impurity of order $V_I=10t_w$. (a) Atomic impurity. (b) Molecular impurity. In both cases, all clusters are affected by its presence.}
    \label{fig:STRONGDIS}
\end{figure}

Based on our numerical implementations, a breakdown of this behavior is observed for strong impurity strengths. This is illustrated in Figs.~\ref{fig:STRONGDIS} (a) and (b), where we plot two graphs with a strong on site impurity placed on an atomic and a molecular site, respectively. 
These observations can be quantitatively substantiated by calculating overlap integrals. For this reason, we will show that in the presence of a weak atomic (molecular) impurity, it is mostly the atomic (molecular) cluster that is affected. while the molecular (atomic) cluster is left intact. In order to see that, we need to calculate cluster-averaged overlap integrals, that is the average of overlap integrals between the clean Fibonacci chain and the one with an impurity. These are given by 
\begin{align}
    \langle O\rangle_A&=\frac{1}{F_{N-3}}\sum_{i=1}^{F_{N-3}}\left(\sum_{j=1}^{F_N}\left|\bra{\psi^0_{i,A}(x_j)}\ket{\psi^I_{i,A}(x_j)}\right|\right), \\
    \langle O\rangle_M&=\frac{1}{2F_{N-2}}\sum_{i=1}^{2F_{N-2}}\left(\sum_{j=1}^{F_N}\left|\bra{\psi^0_{i,M}(x_j)}\ket{\psi^I_{i,M}(x_j)}\right|\right).
\end{align}
\begin{figure}[!hbt]
    \centering
    \includegraphics[width=\columnwidth]{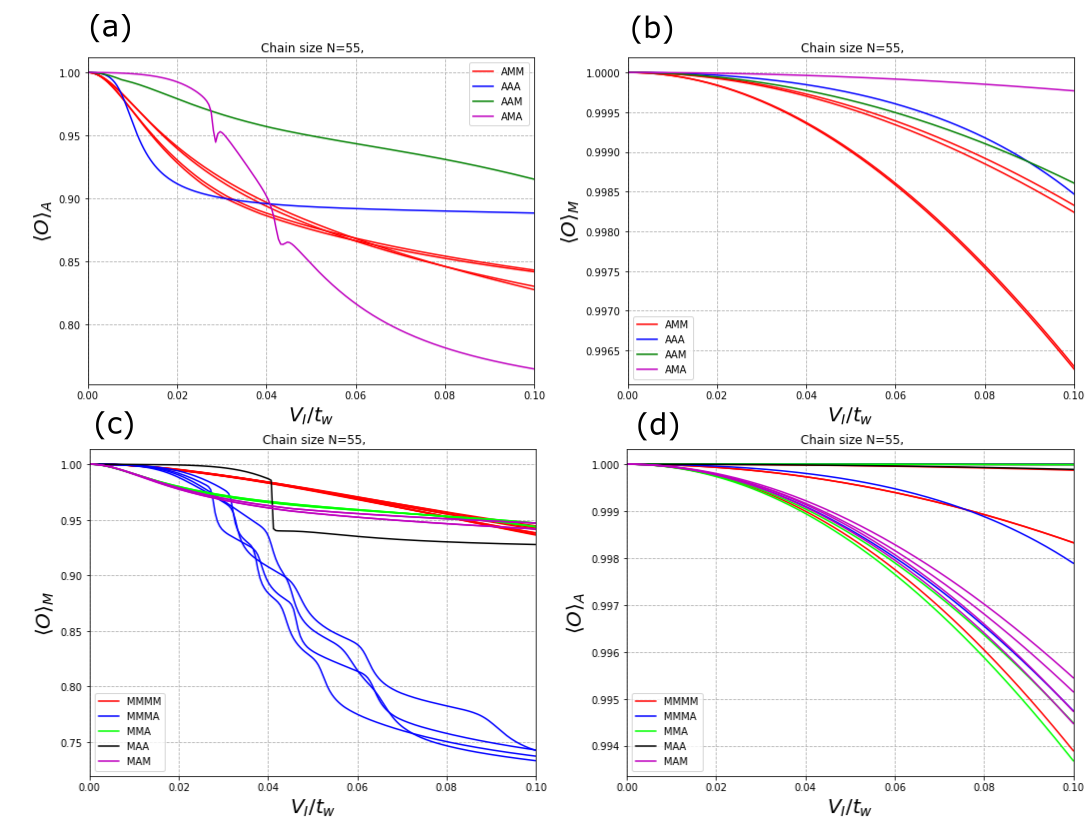}
    \caption{(Color online) Cluster averaged overlap integrals in the presence of (a,b) an atomic impurity and (c,d) a molecular impurity, for all possible impurities that result in different realizations, and which are labelled by the renormalization path of the site at which they are placed. It is clear that the atomic (molecular) cluster is left mostly intact by the presence of a molecular (atomic) impurity for the strengths considered. The parameters considered are still the same as for the previous figures, with $\rho=0.2$ and the number of sites $N=55$.}
    \label{fig:OLPATO}
\end{figure}
\begin{figure}[!hbt]
    \centering
     \includegraphics[width=\columnwidth]{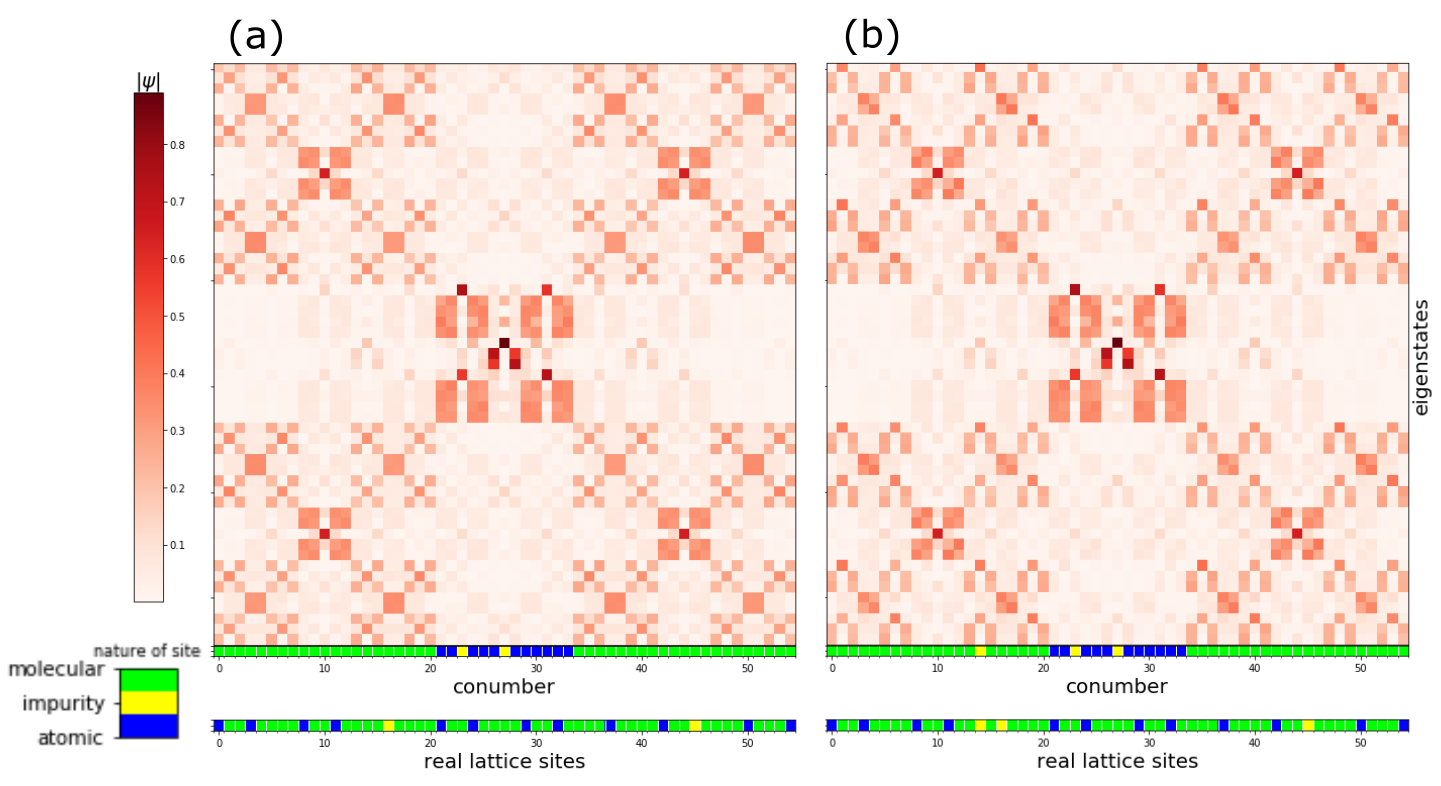}
    \caption{(Color online) Example of multiple weak impurities ($V_I=0.1t_w)$ placed on the Fibonacci chain. (a) AAA-AMA pattern. (b) AAA-AMA-MMMA pattern. It is remarkable that one can just add the individual contributions from each of the classes to generate the (weak) multi-impurity classes.}
    \label{fig:MULTIDISO}
\end{figure}
with $\ket{\psi^0_{i,X}(x_j)}$ and $\ket{\psi^I_{i,X}(x_j)}$ the clean and perturbed eigenstates, respectively, of the atomic/molecular cluster (for $X=A,M$). The results are shown in Fig.~\ref{fig:OLPATO}(a-d), where we see that for the weak impurity strengths considered, the overlaps stay very close to 1 when the the type of the site does not belong to the cluster considered, while it decreases more steeply when the impurity site does belong to the cluster. We have also verified whether these observations are size dependent features, but it turns out not to be the case, as we show in Appendix A.
\begin{figure}[!hbt]
    \centering
    \includegraphics[width=0.85\columnwidth]{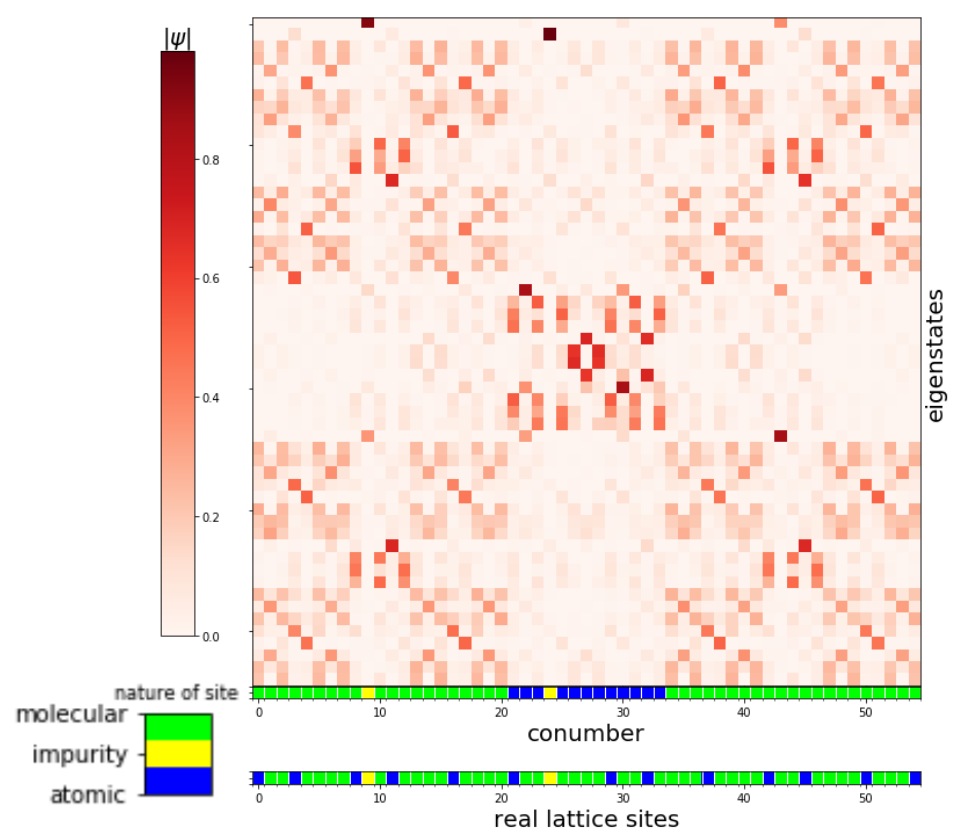}
    \caption{(Color online) Two relatively strong impurities placed on an MAM site and a AMM site. Their strength is set at $V_I=10t_w$. The result is not a superposition of the individual MAM and AMM graphs (see Fig.~\ref{fig:STRONGDIS} for comparison).}
    \label{fig:STRONGMULTIMP}
\end{figure}
\begin{figure*}[!hbt]
    \centering
    \includegraphics[width=0.9\textwidth]{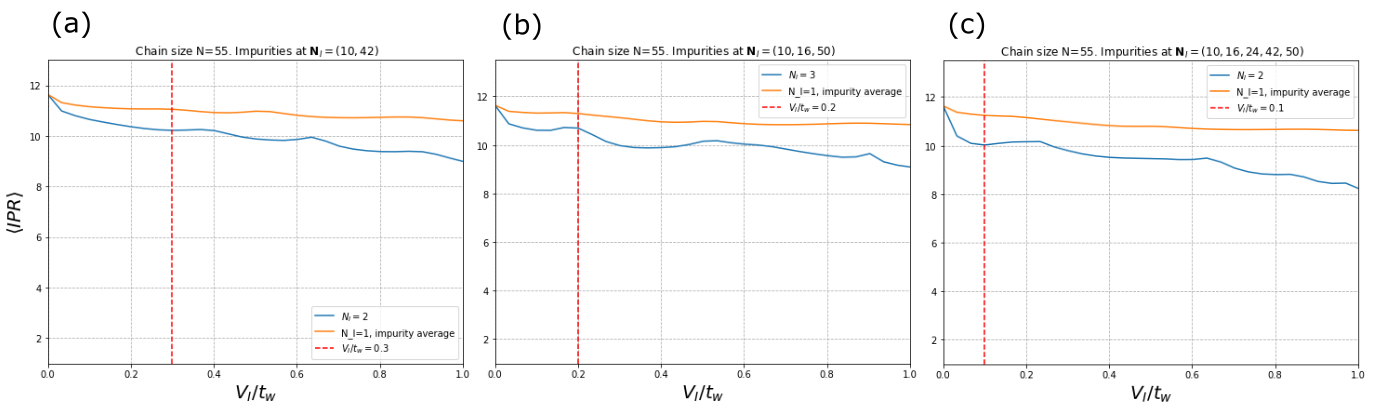}
    \caption{(Color online) Comparisons between the impurity averaged IPRs of single-impurity realizations and the IPR of the multi-impurity system. (a) Two impurities, (b) three impurities and (c) five impurities. We see that they approximately follow a similar evolution until an arbitrarily set threshold (red dashed line), after which the variations seem to not be correlated.}
    \label{fig:IPRMULTIMP}
\end{figure*}
Another interesting property is the additivity of the disrupted patterns, when one adds several weak impurities. It seems that one can just superpose the eigenstate maps of the individual impurity disruptions to obtain the total disordered pattern. This is illustrated by the simple examples in Fig.~\ref{fig:MULTIDISO}: there are two atomic impurities in Fig.~\ref{fig:MULTIDISO}(a), characterized by the renormalization paths AAA+AMA, and in Fig.~\ref{fig:MULTIDISO}(b), we add a molecular impurity to the previous two, such that the total disruption is characterized by the paths AAA+AMA+MMMA. It clearly looks like the graphs have been superimposed on top of one another, as seen from the graphs in Figs.~\ref{fig:classesATO}(b,c) and Fig.~\ref{fig:classesMOL}(c).
Naturally, there is a limitation to these observations, as the amount of paths grows substantially with the size of the chain. Moreover, as the impurities grow stronger, the additivity of the graphs breaks down and the impurities start to influence each other, an effect that is easily visible in the eigenstate map in  Fig.~\ref{fig:STRONGMULTIMP}. 
There we added an impurity with ten times the previous strength on a AAA site and another one on an MMMA site. We can clearly see that the two individual graphs are not superimposed. Additionally, this superposition is not exact even for weak impurities. Indeed, since the eigenstate map represents probability densities, we know that in the case of multiple weak impurities, we can write down the amplitudes to first order in perturbation theory as
\begin{equation}
    \psi_\alpha(x_i)=\alpha(x_i)+V_I\sum_{\beta\neq\alpha}\frac{\beta(x_i)}{E^{(\alpha)}_0-E^{(\beta)}_0}\sum_m\alpha(x_m)\beta^*(x_m)
\end{equation}
where $\alpha$ and $\beta$ denote eigenstates of the pure Fibonacci chain (without impurities). Since we are depicting a conumbered version of $|\psi_\alpha(x_i)|^2$, we also have nonlinearities entering the picture. In the case of very weak impurities, they are overshadowed by the linear term, which is why we observe this additive feature.
The nonlinearities are more pronounced in Fig.~\ref{fig:IPRMULTIMP}, where we plot the impurity-averaged inverse participation ratio (IPR, introduced in Sec.~IIIB) of the single impurity realisations and compare it to the IPR of the system containing all the impurities. We see that they have the same behavior in terms of how they depend on the impurity strength, but they are not equal. Moreover, the similarities seem to quickly disappear with increasing impurity strength, rendering the classification of multiple impurities more complicated than for the single impurity case. The nonlinearities are also more strongly pronounced in this case, which is expected because we are looking at the fourth moment of probability density. 

At this point, we note that the existence of resonant states \cite{subsdisorder}, especially in the case of weak impurities (of order $0.1t_w$), allows for what can be called a ``transition regime'', in which some level of quasiperiodic order is still present in parts of the system. We shall see in the next section that this can also be characterized by studying the localization properties of the system.
\subsection{Localization Properties}
\begin{figure}[!hbt]
    \includegraphics[width=0.9\columnwidth]{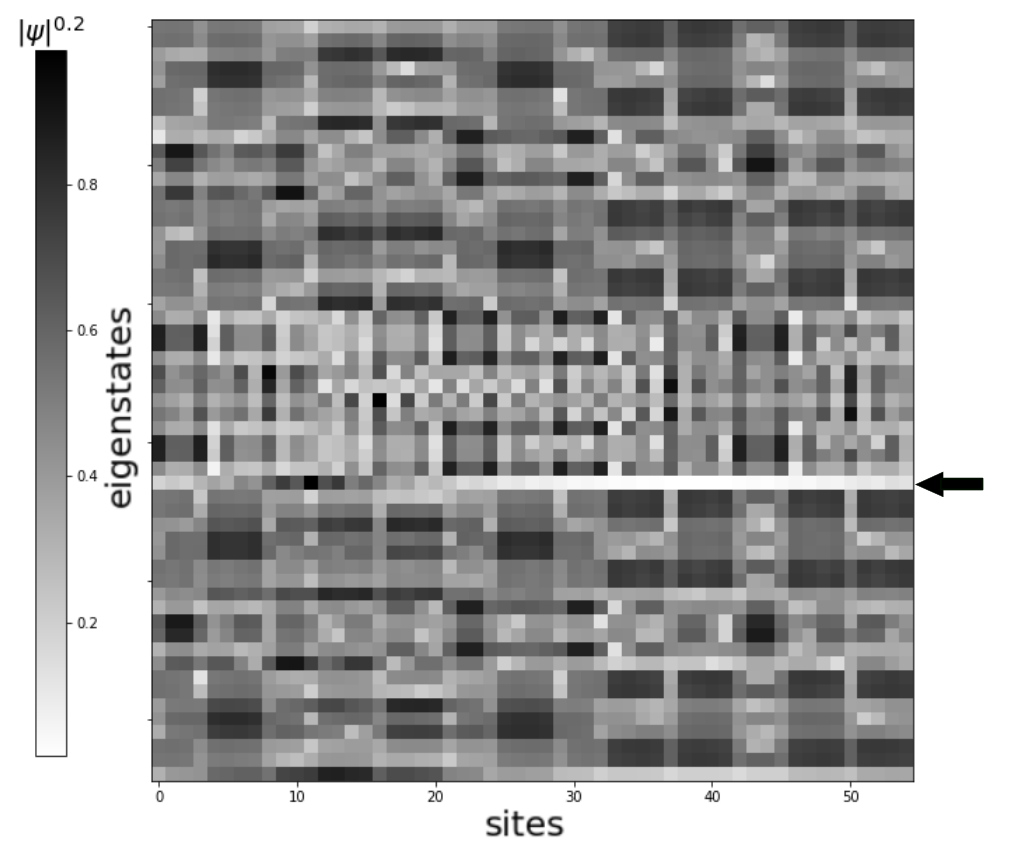}
    \caption{\label{fig:LOCPDENS} Impurity placed on a generation 9 chain (55 sites) at site number 11. The localized state is clearly visible with its white stripe extending over a wide range of sites, as indicated by the black arrow on the right. For the sake of clarity, we used a strong impurity of order 10$t_w$, with the ratio $\rho$=0.2.}
\end{figure}
\begin{figure}
    \centering
    \includegraphics[width=\columnwidth]{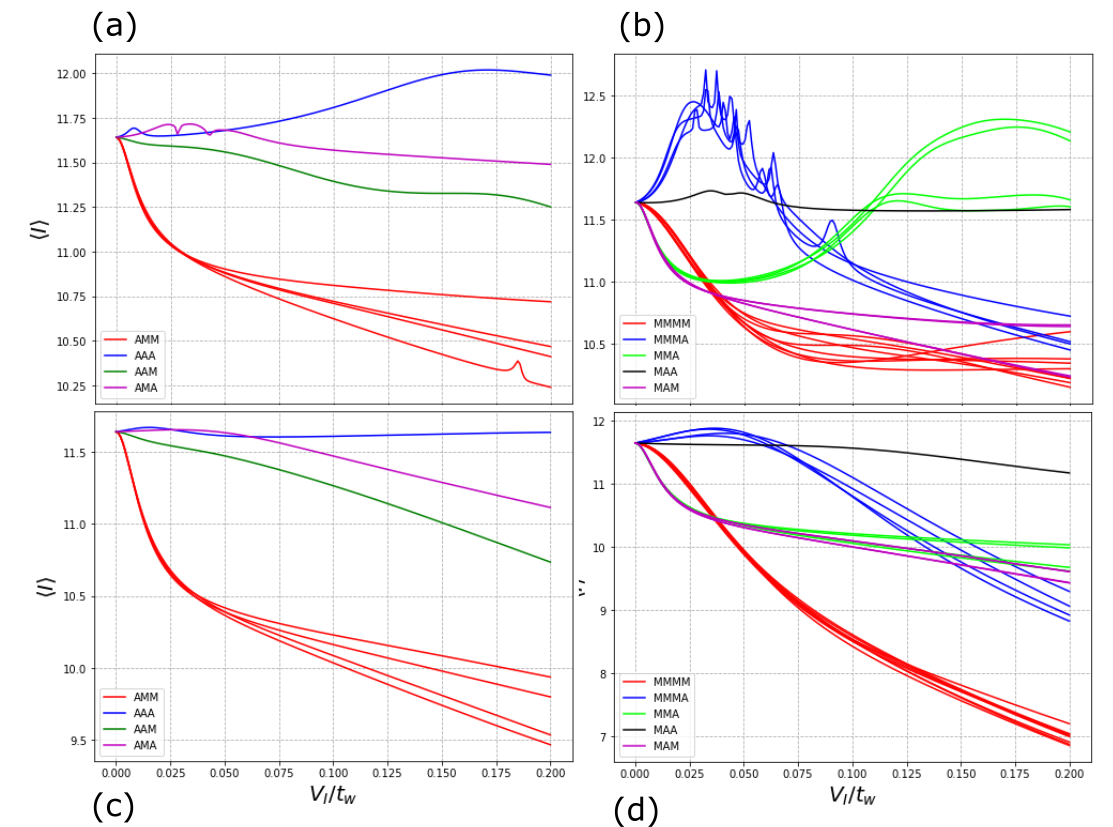}
    \caption{(Color online) (a) Average IPR for the various different realizations of atomic impurity placements. (b) Average IPR for the various realizations of molecular impurity placements. (c,d) show the same quantity as approximated using nondegenerate perturbation theory. It is clear that for weak impurity strengths up to at least $V_I\approx 0.1t_w$, the behaviours are grouped in terms of the nature of the site at which the impurities have been placed. Naturally, we label them by the renormalization paths that they belong to. The system size is again $N=55$ and the modulation strength $\rho=0.2$. We also note that the discrepancies between the approximated and the exact IPRs can be resolved by adding higher order terms in the perturbation theory.}
    \label{fig:AVGIPR}
\end{figure}

We begin by making the general observation that introducing a strong impurity generally leads to the localization of one or several states. This is promptly visualized when one inspects the numerically obtained eigenstate map in Fig. \ref{fig:LOCPDENS}, where the localization can be visualised for one of the eigenstates (white stripe, extending over several sites, below the center of the figure).
This kind of behavior marks a departure from criticality, i.e. the wavefunctions of quasiperiodic systems are neither extended nor localized. In Fig.~\ref{fig:LOCPDENS}, we see that the localization is around the site at which the impurity was placed (site number 11 on the figure, where the darkest square marks the strongest amplitude).  
In order to properly determine the localization behavior of the system, we will calculate the IPR. For a particular eigenstate $\ket{\alpha}=\sum_i\psi^m_\alpha(x_i)\ket{i}$, where proper normalization is imposed ($\sum_i|\psi^m_\alpha(x_i)|^2=1$), it is defined as
\begin{equation}
    I_\alpha=\left(\sum_{j=1}^{F_N}|\psi_{\alpha}(x_j)|^4\right)^{-1}.
\end{equation}
This quantity gives a good measure of localization: in the fully localized limit, where we have $|\psi(x_m)|^2=1$ for some site $x_m$ and zero otherwise, $I_\alpha=1$. In the fully delocalized limit, where the probability density is uniform ($|\psi(x_i)|^2=1/N$ for all sites $x_i$), $I_\alpha=N$. By introducing an impurity, the symmetry between the renormalization paths of energy levels and that of the sites is lost. However, we still want to be able to characterize the localization properties using the renormalization path of the sites. Therefore, instead of looking at each state's IPR, we look at the average IPR over all states. Thus, we define the state-averaged IPR as
\begin{equation}
    \langle I \rangle=\frac{1}{F_{N}}\sum_{\alpha=1}^{F_{N}}I_\alpha=\frac{1}{F_{N}}\sum_{\alpha=1}^{F_{N}}\left(\sum_{j=1}^{F_N}|\psi_{\alpha}(x_j)|^4\right)^{-1}.
\end{equation}
We have plotted various results of $\langle I\rangle$ in Fig.~\ref{fig:AVGIPR}. We observe that, as hinted from the previous section, the behaviour of the IPR can be grouped in terms of the renormalization path that a site belongs to. For the atomic case, all distinct representatives of the four renormalization paths of the $N=55$ chain are plotted (there are seven distinct curves). There is one curve for AAA, AAM and AMA, while there are four curves for AMM. Up to $V_I=0.1t_w$, one observes that curves belonging to the same renormalization path follow the same evolution as a function of impurity strength. In the molecular case, we have again plotted all distinct representatives (19 curves). There are eight curves for MMMM, four for MMMA, four for MMA, two for MAM and one for MAA. They are also relatively well grouped up in terms of renormalization path, up to a strength $V_I=0.1t_w$. 
We can in fact understand why this grouping happens by using first order nondegenerate perturbation theory. The IPR in this case is given by (see App.~\ref{B})
\begin{equation}
     I^m_\alpha=\frac{\sum_i|\psi^m_\alpha(x_i)|^2}{\sum_i|\psi^m_\alpha(x_i)|^4}, 
\end{equation}
where the non-normalized amplitudes, to first order in perturbation theory, are given by
\begin{equation}
    \psi^m_\alpha(x_i)=\alpha(x_i)+V_I\sum_{\beta\neq\alpha}\frac{\alpha(x_m)\beta^*(x_m)\beta(x_i)}{E^{(\alpha)}_0-E^{(\beta)}_0}.
\end{equation}
Here $\ket{\alpha}=\sum_i\alpha(x_i)\ket{i}$ (same for $\beta$) denote the pure, nondegenerate, Fibonacci energy eigenstates, with $E^{(\alpha)}_0$ their energies. From this expression, we are able to see why the grouping in terms of renormalization paths takes place: the amplitudes $\alpha(x_m)$ and $\beta(x_m)$ will be very similar when the site $x_m$ belongs to the same renormalization path. This is understood from the perturbative treatment of the clean Fibonacci chain, where it is known that that to zeroth order, the amplitudes are strictly localized on sites belonging to the same renormalization path \cite{NORI,Mac__2016}. The next orders correct for this strict localization, but the highest weight still remains on the sites belonging to the same renormalization path. This also explains the origin of the symmetric fractal picture that we observed in Sec.~II. For comparison, we plotted the IPR calculated perturbatively in Fig.~\ref{fig:AVGIPR} (c) and (d). We see that globally, it provides a very good approximation of the behavior and the grouping is more pronounced. Another feature that we can observe from the IPRs is that there are some impurity configurations that lead to an average delocalization of the states for a range of impurity strengths, as can be seen for example in Fig.~\ref{fig:AVGIPR} (b), where the blue curve shows an increase in IPR before it decreases. This comes as a surprise, as one expects that the introduction of disorder leads to more localization. Similar results were recently obtained in Ref.~\cite{Anuradha} for a system in which random disorder was introduced in the hopping parameters. 

\section{Conclusion}
To conclude, in this work we studied the effect of disorder by introducing a impurities in the Fibonacci chain. We first introduced the 1D model in the tight-binding approximation and briefly explained how to understand the spectrum through a deflation procedure. This was then followed by the introduction of an impurity in the quasiperiodic lattice, which leads to the appearance of at least one localized state. We have then shown that in the weak impurity regime, disorder is introduced in a very structured manner, following a labelling provided by the renormalization path of the sites at which the impurities have been placed. The disorder is restricted to subclusters of the system. This indicates that there exists a transition regime between a more insulating state where disorder is dominant, and the typical critical states present in a quasiperiodic lattice. We emphasize that the viewpoint offered in this paper stems from the very important notion of conumbers. They turn out to be essential in revealing the structure with which disorder sets in and makes our observations very intuitive. They were further substantiated by studying the behavior of overlap integrals and localization properties through the IPR.

Although the renormalization scheme is only exact in the limit $\rho\ll1$, even $\rho=0.5$ is sufficiently small for the results to hold, and any $\rho\leq0.2$ leads to well separated energy clusters \cite{NORI}. In this sense, the results of our analysis are not exact, but provide a qualitative understanding of the role played by impurities in a Fibonacci hopping model.
Furthermore, the weak impurity regime was studied perturbatively and yielded qualitatively good results. It seems that this works well as long $V_I<\rho$, which forces the impurity to mainly disrupt a particular cluster. 

As an outlook, it would be interesting to study the effect that disorder has in disrupting the local symmetry structure of the system. This is partly because we originally believed that it could play a role in inhibiting the effect of impurities, as was observed in the case of the AAA graph in Fig.~\ref{fig:classesATO}(b). Indeed, this site is the centre of a very large region of palindromic symmetry. However, this naive interpretation was quickly ruled out as the MMA graph in Fig.~\ref{fig:classesMOL}(a) offered very little disruption, even though it had no locally symmetric region surrounding it.  In recent work, the on-site model of the Fibonacci chain has been studied using a \textit{local resonator mode} framework \cite{R_ntgen_2019}. It would be interesting to find an analog version of this framework in the hopping model, which offers an intuitive understanding of the effect of impurities on local symmetries. This could be helpful to further understand the topological features of the Fibonacci chain. Indeed, it is possible to study the robustness of the topological phase by subjecting the system to impurities. Since the gap labelling theorem has been reinterpreted in terms of renormalization paths \cite{Mac__2016}, the presence of stable gap states can be analyzed by the amount of disorder in the graphs. This in turn should be related to how the impurity breaks the (local) palindromic symmetry or preserves it. Hence, one would be able to evaluate whether the topological phase is protected by this symmetry, or if it is of a different nature.

\begin{acknowledgments}
We would like to thank P. Schmelcher, C. Morfonios and M. Rontgen, with whom we had interesting discussions on the role played by local symmetries in 1D quasiperiodic systems). This work was supported by the NWO (Netherlands Organisation for Scientific Research/ Nederlandse Organisatie voor Wetenschappelijk Onderzoek).
\end{acknowledgments}

\bibliography{MAINnewnew}% Produces the bibliography via BibTeX.

\providecommand{\noopsort}[1]{}\providecommand{\singleletter}[1]{#1}%
\begin{thebibliography}{26}%
\makeatletter
\providecommand \@ifxundefined [1]{%
 \@ifx{#1\undefined}
}%
\providecommand \@ifnum [1]{%
 \ifnum #1\expandafter \@firstoftwo
 \else \expandafter \@secondoftwo
 \fi
}%
\providecommand \@ifx [1]{%
 \ifx #1\expandafter \@firstoftwo
 \else \expandafter \@secondoftwo
 \fi
}%
\providecommand \natexlab [1]{#1}%
\providecommand \enquote  [1]{``#1''}%
\providecommand \bibnamefont  [1]{#1}%
\providecommand \bibfnamefont [1]{#1}%
\providecommand \citenamefont [1]{#1}%
\providecommand \href@noop [0]{\@secondoftwo}%
\providecommand \href [0]{\begingroup \@sanitize@url \@href}%
\providecommand \@href[1]{\@@startlink{#1}\@@href}%
\providecommand \@@href[1]{\endgroup#1\@@endlink}%
\providecommand \@sanitize@url [0]{\catcode `\\12\catcode `\$12\catcode
  `\&12\catcode `\#12\catcode `\^12\catcode `\_12\catcode `\%12\relax}%
\providecommand \@@startlink[1]{}%
\providecommand \@@endlink[0]{}%
\providecommand \url  [0]{\begingroup\@sanitize@url \@url }%
\providecommand \@url [1]{\endgroup\@href {#1}{\urlprefix }}%
\providecommand \urlprefix  [0]{URL }%
\providecommand \Eprint [0]{\href }%
\providecommand \doibase [0]{https://doi.org/}%
\providecommand \selectlanguage [0]{\@gobble}%
\providecommand \bibinfo  [0]{\@secondoftwo}%
\providecommand \bibfield  [0]{\@secondoftwo}%
\providecommand \translation [1]{[#1]}%
\providecommand \BibitemOpen [0]{}%
\providecommand \bibitemStop [0]{}%
\providecommand \bibitemNoStop [0]{.\EOS\space}%
\providecommand \EOS [0]{\spacefactor3000\relax}%
\providecommand \BibitemShut  [1]{\csname bibitem#1\endcsname}%
\let\auto@bib@innerbib\@empty
%</preamble>
\bibitem [{\citenamefont {Shechtman}\ \emph {et~al.}(1984)\citenamefont
  {Shechtman}, \citenamefont {Blech}, \citenamefont {Gratias},\ and\
  \citenamefont {Cahn}}]{shechtman}%
  \BibitemOpen
  \bibfield  {author} {\bibinfo {author} {\bibfnamefont {D.}~\bibnamefont
  {Shechtman}}, \bibinfo {author} {\bibfnamefont {I.}~\bibnamefont {Blech}},
  \bibinfo {author} {\bibfnamefont {D.}~\bibnamefont {Gratias}},\ and\ \bibinfo
  {author} {\bibfnamefont {J.~W.}\ \bibnamefont {Cahn}},\ }\bibfield  {title}
  {\bibinfo {title} {Metallic phase with long-range orientational order and no
  translational symmetry},\ }\href
  {https://doi.org/10.1103/PhysRevLett.53.1951} {\bibfield  {journal} {\bibinfo
   {journal} {Phys. Rev. Lett.}\ }\textbf {\bibinfo {volume} {53}},\ \bibinfo
  {pages} {1951} (\bibinfo {year} {1984})}\BibitemShut {NoStop}%
\bibitem [{\citenamefont {Pope}\ and\ \citenamefont
  {Tritt}(2004)}]{popeconduc}%
  \BibitemOpen
  \bibfield  {author} {\bibinfo {author} {\bibfnamefont {A.~L.}\ \bibnamefont
  {Pope}}\ and\ \bibinfo {author} {\bibfnamefont {T.}~\bibnamefont {Tritt}},\
  }\bibinfo {title} {Thermal conductivity of quasicrystalline materials},\ in\
  \href {https://doi.org/10.1007/0-387-26017-X_11} {\emph {\bibinfo {booktitle}
  {Thermal Conductivity: Theory, Properties, and Applications}}},\ \bibinfo
  {editor} {edited by\ \bibinfo {editor} {\bibfnamefont {T.}~\bibnamefont
  {Tritt}}}\ (\bibinfo  {publisher} {Springer US},\ \bibinfo {address} {Boston,
  MA},\ \bibinfo {year} {2004})\ pp.\ \bibinfo {pages} {255--259}\BibitemShut
  {NoStop}%
\bibitem [{\citenamefont {Berger}\ \emph {et~al.}(1993)\citenamefont {Berger},
  \citenamefont {Grenet}, \citenamefont {Lindqvist}, \citenamefont {Lanco},
  \citenamefont {Grieco}, \citenamefont {Fourcaudot},\ and\ \citenamefont
  {Cyrot-Lackmann}}]{alpdreelec}%
  \BibitemOpen
  \bibfield  {author} {\bibinfo {author} {\bibfnamefont {C.}~\bibnamefont
  {Berger}}, \bibinfo {author} {\bibfnamefont {T.}~\bibnamefont {Grenet}},
  \bibinfo {author} {\bibfnamefont {P.}~\bibnamefont {Lindqvist}}, \bibinfo
  {author} {\bibfnamefont {P.}~\bibnamefont {Lanco}}, \bibinfo {author}
  {\bibfnamefont {J.}~\bibnamefont {Grieco}}, \bibinfo {author} {\bibfnamefont
  {G.}~\bibnamefont {Fourcaudot}},\ and\ \bibinfo {author} {\bibfnamefont
  {F.}~\bibnamefont {Cyrot-Lackmann}},\ }\bibfield  {title} {\bibinfo {title}
  {The new alpdre icosahedral phase: Towards universal electronic behaviour for
  quasicrystals?},\ }\href
  {https://doi.org/https://doi.org/10.1016/0038-1098(93)90543-V} {\bibfield
  {journal} {\bibinfo  {journal} {Solid State Communications}\ }\textbf
  {\bibinfo {volume} {87}},\ \bibinfo {pages} {977 } (\bibinfo {year}
  {1993})}\BibitemShut {NoStop}%
\bibitem [{\citenamefont {Dubois}(2002)}]{quasifeaturesbook}%
  \BibitemOpen
  \bibfield  {author} {\bibinfo {author} {\bibfnamefont {J.}~\bibnamefont
  {Dubois}},\ }\bibinfo {title} {Bulk and surface properties of
  quasicrystalline materials and their potential applications},\ in\ \href
  {https://doi.org/10.1007/978-3-662-05028-6_26} {\emph {\bibinfo {booktitle}
  {Quasicrystals, An Introduction to Structure, Physical Properties and
  Applications}}}\ (\bibinfo {year} {2002})\ pp.\ \bibinfo {pages}
  {507--538}\BibitemShut {NoStop}%
\bibitem [{\citenamefont {Maciá}(2005)}]{emaciaaporder}%
  \BibitemOpen
  \bibfield  {author} {\bibinfo {author} {\bibfnamefont {E.}~\bibnamefont
  {Maciá}},\ }\bibfield  {title} {\bibinfo {title} {The role of aperiodic
  order in science and technology},\ }\href
  {https://doi.org/10.1088/0034-4885/69/2/r03} {\bibfield  {journal} {\bibinfo
  {journal} {Reports on Progress in Physics}\ }\textbf {\bibinfo {volume}
  {69}},\ \bibinfo {pages} {397} (\bibinfo {year} {2005})}\BibitemShut
  {NoStop}%
\bibitem [{\citenamefont {{de Bruijn}}(1981)}]{DEBRUIJN198153}%
  \BibitemOpen
  \bibfield  {author} {\bibinfo {author} {\bibfnamefont {N.}~\bibnamefont {{de
  Bruijn}}},\ }\bibfield  {title} {\bibinfo {title} {Algebraic theory of
  penrose's non-periodic tilings of the plane. ii},\ }\href
  {https://doi.org/https://doi.org/10.1016/1385-7258(81)90017-2} {\bibfield
  {journal} {\bibinfo  {journal} {Indagationes Mathematicae (Proceedings)}\
  }\textbf {\bibinfo {volume} {84}},\ \bibinfo {pages} {53 } (\bibinfo {year}
  {1981})}\BibitemShut {NoStop}%
\bibitem [{\citenamefont {de~Boissieu}(2019)}]{tedjanssen}%
  \BibitemOpen
  \bibfield  {author} {\bibinfo {author} {\bibfnamefont {M.}~\bibnamefont
  {de~Boissieu}},\ }\bibfield  {title} {\bibinfo {title} {{Ted {J}anssen and
  aperiodic crystals}},\ }\href {https://doi.org/doi:10.1107/S2053273318016765}
  {\bibfield  {journal} {\bibinfo  {journal} {Act. Crystall. Sect. A}\ }\textbf
  {\bibinfo {volume} {75(Pt 2)}},\ \bibinfo {pages} {273–280.} (\bibinfo
  {year} {2019})}\BibitemShut {NoStop}%
\bibitem [{\citenamefont {Janssen}(1986)}]{Janssen:a25379}%
  \BibitemOpen
  \bibfield  {author} {\bibinfo {author} {\bibfnamefont {T.}~\bibnamefont
  {Janssen}},\ }\bibfield  {title} {\bibinfo {title} {{Crystallography of
  quasi-crystals}},\ }\href {https://doi.org/10.1107/S0108767386099324}
  {\bibfield  {journal} {\bibinfo  {journal} {Act. Crystall. Sect. A}\ }\textbf
  {\bibinfo {volume} {42}},\ \bibinfo {pages} {261} (\bibinfo {year}
  {1986})}\BibitemShut {NoStop}%
\bibitem [{\citenamefont {Bindi}\ \emph {et~al.}(2009)\citenamefont {Bindi},
  \citenamefont {Steinhardt}, \citenamefont {Yao},\ and\ \citenamefont
  {Lu}}]{Bindi1306}%
  \BibitemOpen
  \bibfield  {author} {\bibinfo {author} {\bibfnamefont {L.}~\bibnamefont
  {Bindi}}, \bibinfo {author} {\bibfnamefont {P.~J.}\ \bibnamefont
  {Steinhardt}}, \bibinfo {author} {\bibfnamefont {N.}~\bibnamefont {Yao}},\
  and\ \bibinfo {author} {\bibfnamefont {P.~J.}\ \bibnamefont {Lu}},\
  }\bibfield  {title} {\bibinfo {title} {Natural {Q}uasicrystals},\ }\href
  {https://doi.org/10.1126/science.1170827} {\bibfield  {journal} {\bibinfo
  {journal} {Science}\ }\textbf {\bibinfo {volume} {324}},\ \bibinfo {pages}
  {1306} (\bibinfo {year} {2009})}\BibitemShut {NoStop}%
\bibitem [{\citenamefont {Flicker}\ and\ \citenamefont {van
  Wezel}(2015)}]{CDWQUASIPER}%
  \BibitemOpen
  \bibfield  {author} {\bibinfo {author} {\bibfnamefont {F.}~\bibnamefont
  {Flicker}}\ and\ \bibinfo {author} {\bibfnamefont {J.}~\bibnamefont {van
  Wezel}},\ }\bibfield  {title} {\bibinfo {title} {One-dimensional
  quasicrystals from incommensurate charge order},\ }\href
  {https://doi.org/10.1103/PhysRevLett.115.236401} {\bibfield  {journal}
  {\bibinfo  {journal} {Phys. Rev. Lett.}\ }\textbf {\bibinfo {volume} {115}},\
  \bibinfo {pages} {236401} (\bibinfo {year} {2015})}\BibitemShut {NoStop}%
\bibitem [{\citenamefont {Niu}\ and\ \citenamefont {Nori}(1991)}]{NORI}%
  \BibitemOpen
  \bibfield  {author} {\bibinfo {author} {\bibfnamefont {Q.}~\bibnamefont
  {Niu}}\ and\ \bibinfo {author} {\bibfnamefont {F.}~\bibnamefont {Nori}},\
  }\bibfield  {title} {\bibinfo {title} {Spectral splitting and wave-function
  scaling in quasicrystalline and hierarchical structures},\ }\href
  {https://doi.org/10.1103/PhysRevB.42.10329} {\bibfield  {journal} {\bibinfo
  {journal} {Phys. Rev. B}\ }\textbf {\bibinfo {volume} {42}},\ \bibinfo
  {pages} {10329} (\bibinfo {year} {1991})}\BibitemShut {NoStop}%
\bibitem [{\citenamefont {Kohmoto}\ and\ \citenamefont {Banavar}(1986)}]{KOHO}%
  \BibitemOpen
  \bibfield  {author} {\bibinfo {author} {\bibfnamefont {M.}~\bibnamefont
  {Kohmoto}}\ and\ \bibinfo {author} {\bibfnamefont {J.}~\bibnamefont
  {Banavar}},\ }\bibfield  {title} {\bibinfo {title} {Quasiperiodic lattice:
  Electronic properties, phonon properties, and diffusion},\ }\href
  {https://doi.org/10.1103/PhysRevB.34.563} {\bibfield  {journal} {\bibinfo
  {journal} {Phys. Rev. B}\ }\textbf {\bibinfo {volume} {34}},\ \bibinfo
  {pages} {563} (\bibinfo {year} {1986})}\BibitemShut {NoStop}%
\bibitem [{\citenamefont {Bellissard}(1992)}]{Bellissard1992}%
  \BibitemOpen
  \bibfield  {author} {\bibinfo {author} {\bibfnamefont {J.}~\bibnamefont
  {Bellissard}},\ }\bibinfo {title} {Gap labelling theorems for schr{\"o}dinger
  operators},\ in\ \href {https://doi.org/10.1007/978-3-662-02838-4_12} {\emph
  {\bibinfo {booktitle} {From Number Theory to Physics}}},\ \bibinfo {editor}
  {edited by\ \bibinfo {editor} {\bibfnamefont {M.}~\bibnamefont
  {Waldschmidt}}, \bibinfo {editor} {\bibfnamefont {P.}~\bibnamefont {Moussa}},
  \bibinfo {editor} {\bibfnamefont {J.-M.}\ \bibnamefont {Luck}},\ and\
  \bibinfo {editor} {\bibfnamefont {C.}~\bibnamefont {Itzykson}}}\ (\bibinfo
  {publisher} {Springer Berlin Heidelberg},\ \bibinfo {address} {Berlin,
  Heidelberg},\ \bibinfo {year} {1992})\ pp.\ \bibinfo {pages}
  {538--630}\BibitemShut {NoStop}%
\bibitem [{\citenamefont {Sire}\ and\ \citenamefont
  {Mosseri}(1990)}]{SireMosseri}%
  \BibitemOpen
  \bibfield  {author} {\bibinfo {author} {\bibfnamefont {C.}~\bibnamefont
  {Sire}}\ and\ \bibinfo {author} {\bibfnamefont {R.}~\bibnamefont {Mosseri}},\
  }\bibfield  {title} {\bibinfo {title} {Excitation spectrum, extended states,
  gap closing : some exact results for codimension one quasicrystals},\ }\href
  {https://doi.org/10.1051/jphys:0199000510150156900} {\bibfield  {journal}
  {\bibinfo  {journal} {Journal de Physique}\ }\textbf {\bibinfo {volume}
  {51}},\ \bibinfo {pages} {1569 } (\bibinfo {year} {1990})}\BibitemShut
  {NoStop}%
\bibitem [{\citenamefont {Mac\'e}\ \emph {et~al.}(2016)\citenamefont {Mac\'e},
  \citenamefont {Jagannathan},\ and\ \citenamefont {Pi\'echon}}]{Mac__2016}%
  \BibitemOpen
  \bibfield  {author} {\bibinfo {author} {\bibfnamefont {N.}~\bibnamefont
  {Mac\'e}}, \bibinfo {author} {\bibfnamefont {A.}~\bibnamefont
  {Jagannathan}},\ and\ \bibinfo {author} {\bibfnamefont {F.}~\bibnamefont
  {Pi\'echon}},\ }\bibfield  {title} {\bibinfo {title} {Fractal dimensions of
  wave functions and local spectral measures on the {F}ibonacci chain},\ }\href
  {https://doi.org/10.1103/PhysRevB.93.205153} {\bibfield  {journal} {\bibinfo
  {journal} {Phys. Rev. B}\ }\textbf {\bibinfo {volume} {93}},\ \bibinfo
  {pages} {205153} (\bibinfo {year} {2016})}\BibitemShut {NoStop}%
\bibitem [{\citenamefont {Levy}\ \emph {et~al.}(2016)\citenamefont {Levy},
  \citenamefont {Barak}, \citenamefont {Fisher},\ and\ \citenamefont
  {Akkermans}}]{levy2016topological}%
  \BibitemOpen
  \bibfield  {author} {\bibinfo {author} {\bibfnamefont {E.}~\bibnamefont
  {Levy}}, \bibinfo {author} {\bibfnamefont {A.}~\bibnamefont {Barak}},
  \bibinfo {author} {\bibfnamefont {A.}~\bibnamefont {Fisher}},\ and\ \bibinfo
  {author} {\bibfnamefont {E.}~\bibnamefont {Akkermans}},\ }\href@noop {}
  {\bibinfo {title} {Topological properties of {F}ibonacci quasicrystals : A
  scattering analysis of {C}hern numbers}} (\bibinfo {year} {2016}),\ \Eprint
  {https://arxiv.org/abs/1509.04028} {arXiv:1509.04028 [physics.optics]}
  \BibitemShut {NoStop}%
\bibitem [{\citenamefont {R\"ontgen}\ \emph {et~al.}(2019)\citenamefont
  {R\"ontgen}, \citenamefont {Morfonios}, \citenamefont {Wang}, \citenamefont
  {Dal~Negro},\ and\ \citenamefont {Schmelcher}}]{R_ntgen_2019}%
  \BibitemOpen
  \bibfield  {author} {\bibinfo {author} {\bibfnamefont {M.}~\bibnamefont
  {R\"ontgen}}, \bibinfo {author} {\bibfnamefont {C.~V.}\ \bibnamefont
  {Morfonios}}, \bibinfo {author} {\bibfnamefont {R.}~\bibnamefont {Wang}},
  \bibinfo {author} {\bibfnamefont {L.}~\bibnamefont {Dal~Negro}},\ and\
  \bibinfo {author} {\bibfnamefont {P.}~\bibnamefont {Schmelcher}},\ }\bibfield
   {title} {\bibinfo {title} {Local symmetry theory of resonator structures for
  the real-space control of edge states in binary aperiodic chains},\ }\href
  {https://doi.org/10.1103/PhysRevB.99.214201} {\bibfield  {journal} {\bibinfo
  {journal} {Phys. Rev. B}\ }\textbf {\bibinfo {volume} {99}},\ \bibinfo
  {pages} {214201} (\bibinfo {year} {2019})}\BibitemShut {NoStop}%
\bibitem [{\citenamefont {Kraus}\ and\ \citenamefont
  {Zilberberg}(2012)}]{Kraus_2012}%
  \BibitemOpen
  \bibfield  {author} {\bibinfo {author} {\bibfnamefont {Y.~E.}\ \bibnamefont
  {Kraus}}\ and\ \bibinfo {author} {\bibfnamefont {O.}~\bibnamefont
  {Zilberberg}},\ }\bibfield  {title} {\bibinfo {title} {Topological
  equivalence between the fibonacci quasicrystal and the harper model},\ }\href
  {https://doi.org/10.1103/PhysRevLett.109.116404} {\bibfield  {journal}
  {\bibinfo  {journal} {Phys. Rev. Lett.}\ }\textbf {\bibinfo {volume} {109}},\
  \bibinfo {pages} {116404} (\bibinfo {year} {2012})}\BibitemShut {NoStop}%
\bibitem [{\citenamefont {Kraus}\ \emph {et~al.}(2012)\citenamefont {Kraus},
  \citenamefont {Lahini}, \citenamefont {Ringel}, \citenamefont {Verbin},\ and\
  \citenamefont {Zilberberg}}]{krausexp}%
  \BibitemOpen
  \bibfield  {author} {\bibinfo {author} {\bibfnamefont {Y.~E.}\ \bibnamefont
  {Kraus}}, \bibinfo {author} {\bibfnamefont {Y.}~\bibnamefont {Lahini}},
  \bibinfo {author} {\bibfnamefont {Z.}~\bibnamefont {Ringel}}, \bibinfo
  {author} {\bibfnamefont {M.}~\bibnamefont {Verbin}},\ and\ \bibinfo {author}
  {\bibfnamefont {O.}~\bibnamefont {Zilberberg}},\ }\bibfield  {title}
  {\bibinfo {title} {Topological states and adiabatic pumping in
  quasicrystals},\ }\href {https://doi.org/10.1103/PhysRevLett.109.106402}
  {\bibfield  {journal} {\bibinfo  {journal} {Phys. Rev. Lett.}\ }\textbf
  {\bibinfo {volume} {109}},\ \bibinfo {pages} {106402} (\bibinfo {year}
  {2012})}\BibitemShut {NoStop}%
\bibitem [{\citenamefont {Rai}\ \emph {et~al.}(2019)\citenamefont {Rai},
  \citenamefont {Haas},\ and\ \citenamefont
  {Jagannathan}}]{PhysRevB.100.165121}%
  \BibitemOpen
  \bibfield  {author} {\bibinfo {author} {\bibfnamefont {G.}~\bibnamefont
  {Rai}}, \bibinfo {author} {\bibfnamefont {S.}~\bibnamefont {Haas}},\ and\
  \bibinfo {author} {\bibfnamefont {A.}~\bibnamefont {Jagannathan}},\
  }\bibfield  {title} {\bibinfo {title} {Proximity effect in a
  superconductor-quasicrystal hybrid ring},\ }\href@noop {} {\bibfield
  {journal} {\bibinfo  {journal} {Phys. Rev. B}\ }\textbf {\bibinfo {volume}
  {100}},\ \bibinfo {pages} {165121} (\bibinfo {year} {2019})}\BibitemShut
  {NoStop}%
\bibitem [{\citenamefont {Rai}\ \emph {et~al.}(2020)\citenamefont {Rai},
  \citenamefont {Haas},\ and\ \citenamefont {Jagannathan}}]{SupCondOPfluc}%
  \BibitemOpen
  \bibfield  {author} {\bibinfo {author} {\bibfnamefont {G.}~\bibnamefont
  {Rai}}, \bibinfo {author} {\bibfnamefont {S.}~\bibnamefont {Haas}},\ and\
  \bibinfo {author} {\bibfnamefont {A.}~\bibnamefont {Jagannathan}},\
  }\bibfield  {title} {\bibinfo {title} {Superconducting proximity effect and
  order parameter fluctuations in disordered and quasiperiodic systems},\
  }\href {https://doi.org/10.1103/PhysRevB.102.134211} {\bibfield  {journal}
  {\bibinfo  {journal} {Phys. Rev. B}\ }\textbf {\bibinfo {volume} {102}},\
  \bibinfo {pages} {134211} (\bibinfo {year} {2020})}\BibitemShut {NoStop}%
\bibitem [{\citenamefont {Naumis}\ and\ \citenamefont
  {Arag\'on}(1996)}]{subsdisorder}%
  \BibitemOpen
  \bibfield  {author} {\bibinfo {author} {\bibfnamefont {G.~G.}\ \bibnamefont
  {Naumis}}\ and\ \bibinfo {author} {\bibfnamefont {J.~L.}\ \bibnamefont
  {Arag\'on}},\ }\bibfield  {title} {\bibinfo {title} {Substitutional disorder
  in a {F}ibonacci chain: Resonant eigenstates and instability of the
  spectrum},\ }\href {https://doi.org/10.1103/PhysRevB.54.15079} {\bibfield
  {journal} {\bibinfo  {journal} {Phys. Rev. B}\ }\textbf {\bibinfo {volume}
  {54}},\ \bibinfo {pages} {15079} (\bibinfo {year} {1996})}\BibitemShut
  {NoStop}%
\bibitem [{\citenamefont {Jagannathan}\ and\ \citenamefont
  {Tarzia}(2020)}]{Anuradha}%
  \BibitemOpen
  \bibfield  {author} {\bibinfo {author} {\bibfnamefont {A.}~\bibnamefont
  {Jagannathan}}\ and\ \bibinfo {author} {\bibfnamefont {M.}~\bibnamefont
  {Tarzia}},\ }\bibfield  {title} {\bibinfo {title} {Re-entrance and
  localization phenomena in disordered {F}ibonacci chains},\ }\href@noop {}
  {\bibfield  {journal} {\bibinfo  {journal} {Eur. Phys. J. B}\ }\textbf
  {\bibinfo {volume} {93}} (\bibinfo {year} {2020})}\BibitemShut {NoStop}%
\bibitem [{\citenamefont {Jagannathan}\ \emph {et~al.}(2019)\citenamefont
  {Jagannathan}, \citenamefont {Jeena},\ and\ \citenamefont
  {Tarzia}}]{NonmonotonicXover}%
  \BibitemOpen
  \bibfield  {author} {\bibinfo {author} {\bibfnamefont {A.}~\bibnamefont
  {Jagannathan}}, \bibinfo {author} {\bibfnamefont {P.}~\bibnamefont {Jeena}},\
  and\ \bibinfo {author} {\bibfnamefont {M.}~\bibnamefont {Tarzia}},\
  }\bibfield  {title} {\bibinfo {title} {Nonmonotonic crossover and scaling
  behavior in a disordered one-dimensional quasicrystal},\ }\href
  {https://doi.org/10.1103/PhysRevB.99.054203} {\bibfield  {journal} {\bibinfo
  {journal} {Phys. Rev. B}\ }\textbf {\bibinfo {volume} {99}},\ \bibinfo
  {pages} {054203} (\bibinfo {year} {2019})}\BibitemShut {NoStop}%
\bibitem [{\citenamefont {Pi\'echon}\ \emph {et~al.}(1995)\citenamefont
  {Pi\'echon}, \citenamefont {Benakli},\ and\ \citenamefont
  {Jagannathan}}]{Pichon}%
  \BibitemOpen
  \bibfield  {author} {\bibinfo {author} {\bibfnamefont {F.}~\bibnamefont
  {Pi\'echon}}, \bibinfo {author} {\bibfnamefont {M.}~\bibnamefont {Benakli}},\
  and\ \bibinfo {author} {\bibfnamefont {A.}~\bibnamefont {Jagannathan}},\
  }\bibfield  {title} {\bibinfo {title} {Analytical results for scaling
  properties of the spectrum of the {F}ibonacci chain},\ }\href
  {https://doi.org/10.1103/PhysRevLett.74.5248} {\bibfield  {journal} {\bibinfo
   {journal} {Phys. Rev. Lett.}\ }\textbf {\bibinfo {volume} {74}},\ \bibinfo
  {pages} {5248} (\bibinfo {year} {1995})}\BibitemShut {NoStop}%
\bibitem [{\citenamefont {Halsey}\ \emph {et~al.}(1986)\citenamefont {Halsey},
  \citenamefont {Jensen}, \citenamefont {Kadanoff}, \citenamefont {Procaccia},\
  and\ \citenamefont {Shraiman}}]{fractalmeasures}%
  \BibitemOpen
  \bibfield  {author} {\bibinfo {author} {\bibfnamefont {T.~C.}\ \bibnamefont
  {Halsey}}, \bibinfo {author} {\bibfnamefont {M.~H.}\ \bibnamefont {Jensen}},
  \bibinfo {author} {\bibfnamefont {L.~P.}\ \bibnamefont {Kadanoff}}, \bibinfo
  {author} {\bibfnamefont {I.}~\bibnamefont {Procaccia}},\ and\ \bibinfo
  {author} {\bibfnamefont {B.~I.}\ \bibnamefont {Shraiman}},\ }\bibfield
  {title} {\bibinfo {title} {Fractal measures and their singularities: The
  characterization of strange sets},\ }\href
  {https://doi.org/10.1103/PhysRevA.33.1141} {\bibfield  {journal} {\bibinfo
  {journal} {Phys. Rev. A}\ }\textbf {\bibinfo {volume} {33}},\ \bibinfo
  {pages} {1141} (\bibinfo {year} {1986})}\BibitemShut {NoStop}%
\end{thebibliography}%

% \clearpage

% \onecolumngrid

% \appendix

% \section{Disordered Graphs}

% \begin{figure}[!hbt]
% \subfloat[No impurity]{\includegraphics[width = 0.362\textwidth]{CLASSS/FREE.png}}
% \subfloat[MMMM]{\includegraphics[width = 0.307\textwidth]{CLASSS/MMMM.png}}
% \subfloat[MMMA]{\includegraphics[width = 0.325\textwidth]{CLASSS/MMMA.png}} \\
% \subfloat[MMA]{\includegraphics[width = 0.36\textwidth]{CLASSS/MMA.png}}
% \subfloat[MAM I]{\includegraphics[width = 0.31\textwidth]{CLASSS/MAMX.png}} 
% \subfloat[MAM II]{\includegraphics[width = 0.315\textwidth]{CLASSS/MAMY.png}}
% \end{figure}

% \begin{figure}\ContinuedFloat
% \subfloat[MAA]{\includegraphics[width = 0.37\textwidth]{CLASSS/MAA.png}}
% \subfloat[AMM I]{\includegraphics[width = 0.305\textwidth]{CLASSS/AMMX.png}} 
% \subfloat[AMM II]{\includegraphics[width = 0.325\textwidth]{CLASSS/AMMY.png}} \\
% \subfloat[AMA]{\includegraphics[width = 0.362\textwidth]{CLASSS/AMA.png}}
% \subfloat[AAM]{\includegraphics[width = 0.305\textwidth]{CLASSS/AAM.png}} 
% \subfloat[AAA]{\includegraphics[width = 0.315\textwidth]{CLASSS/AAA.png}} 
% \caption{}
% \label{fig:classexam}
% \end{figure}

\appendix

\section{Finite-size Scaling of IPR and Size Dependence of Overlap Integrals}\label{A}
\begin{figure}[!hbt]
    \centering
    \includegraphics[width=\columnwidth]{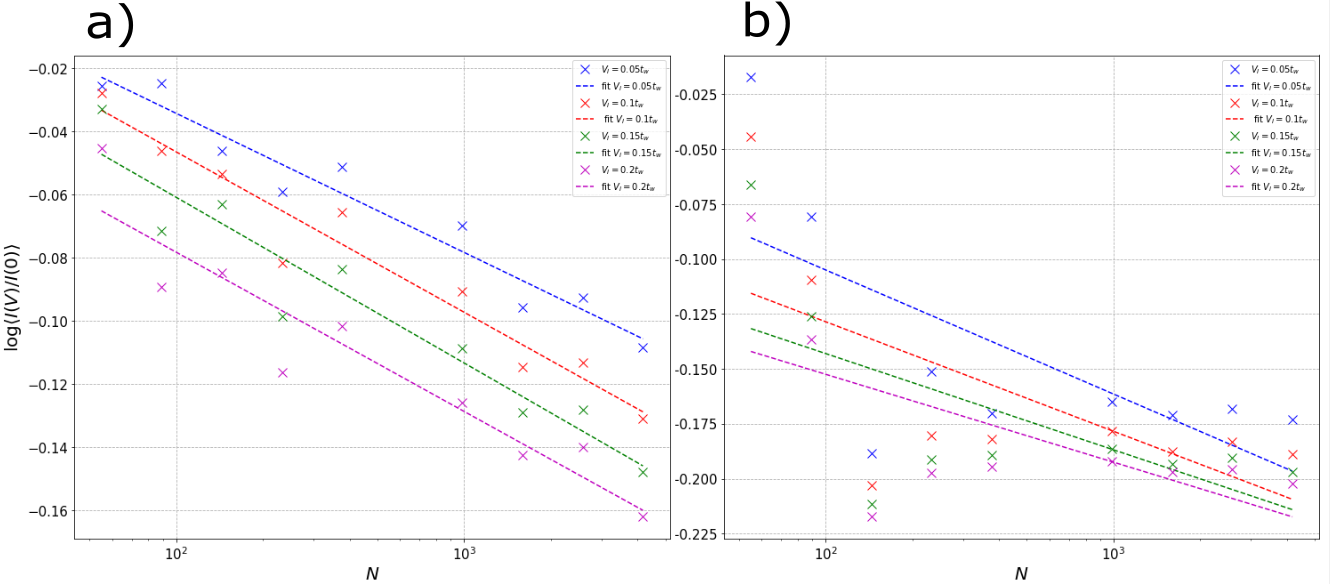}
    \caption{(Color online) Finite size scaling of the averaged IPR in the presence of (a) an atomic impurity and (b) a molecular impurity. In both cases, we have a negative slope in the log-log line, meaning we are witnessing more localization when an impurity is introduced in the system.}
    \label{fig:IPRFSS}
\end{figure}
\begin{figure*}[!hbt]
    \centering
    \includegraphics[width=\textwidth]{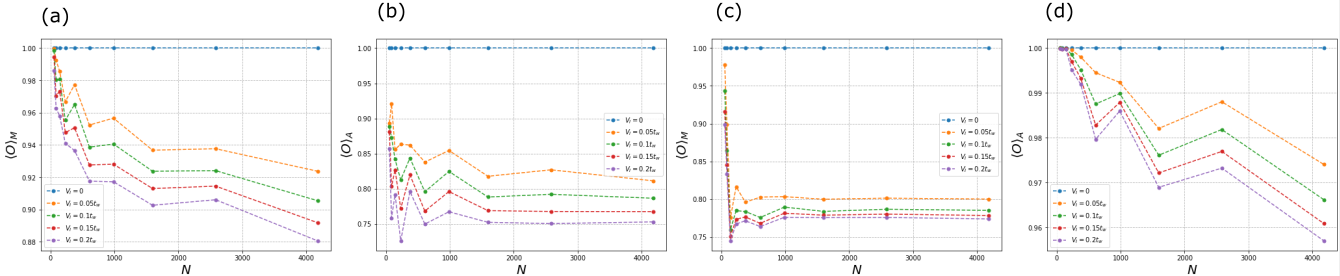}
    \caption{(Color online) Cluster-averaged overlap integral versus system size for eight different system sizes, corresponding to Fibonacci chain approximants ranging from generation 8 to 17. (a,b) Overlap integrals in the presence of an atomic impurity. (c,d) Overlap integrals in the presence of a molecular impurity. In both of them, the cluster that is least disturbed is the one belonging to sites of a different nature than the one on which the impurity has been placed.}
    \label{fig:OLPSIZE}
\end{figure*}
In this section, we provide a finite size scaling analysis plot in Fig.~\ref{fig:IPRFSS} to show that the presence of an impurity induces localization. Since the IPR grows with size as $I_\alpha\propto L^\nu$, with $\nu\to0$ meaning complete localization and $\nu\to1$ full delocalization, as $N\to\infty$, the finite-size scaling analysis is done by numerically calculating the ratio $\langle I(V)\rangle/\langle I(0)\rangle\propto L^\gamma.$
If the log-log plot has a positive slope ($\gamma>0)$, it means the state became more delocalized, and if the slope is negative ($\gamma<0)$, it became more localized in the presence of an impurity. We clearly observe a negative slope in both Figs.~\ref{fig:IPRFSS} (a) and (b), which means that the impurity induces a localization of states, on average.

Furthermore, we show that the general features of the cluster-averaged overlap integrals are not size dependent. In other words, we show that if an impurity has been placed on an atomic site, then the molecular cluster is less affected by its presence than the atomic cluster (vice-versa for a molecular impurity). To this end, we computed the cluster-averaged overlap integral for different Fibonacci approximant chain sizes, choosing just one impurity to represent the general features. The results are plotted in Fig.~\ref{fig:OLPSIZE}. We see that for both cases of atomic and molecular impurities, the behavior described in the bulk of the text is still observed. That is, when an impurity is placed in an atomic (molecular) site, the cluster-averaged overlap integral is larger for the molecular (atomic) cluster, even though in both cases the overlap globally decreases for increasing sizes. These features get washed off as the impurity strength decreases and the size increases. Therefore, there must exist a crossover size $M(V)$ that increases with decreasing impurity strength. This size yields a threshold to the relevance of the results presented in this paper. 

Given an arbitrary threshold of similarity $\langle O\rangle_X=0.9$ for a cluster of type $X$, we observe the following:
In Fig.~\ref{fig:OLPSIZE} (a), for an atomic impurity, the crossover size for the molecular cluster is reached at $N=4181$ starting at impurity strength $V_I=0.15t_w$. In Fig.~\ref{fig:OLPSIZE} (d), for a molecular impurity, we do not reach the crossover size for the atomic cluster for any of the strengths considered, and $\langle O\rangle_A>0.95$ for all sizes considered. In Figs.~\ref{fig:OLPSIZE} (b) and (c), when the cluster type is the same as the impurity type, the threshold is reached very quickly, as expected.

\section{Perturbative calculation of the IPR}\label{B}

Let us consider a system described by the Hamiltonian $H=H_0+V$, where $H_0$ is the usual Fibonacci Hamiltonian and $V$ is the perturbation containing the impurity of strength $V_I$, placed at some site $m$. An eigenstate $\ket{\psi}$ of the Hamiltonian $H$ can be written, to first order in perturbation theory, as
\begin{equation}
    \ket{\psi}=\ket{\alpha}+\sum_{\beta\neq\alpha}\frac{\bra{\beta}V\ket{\alpha}}{E^{(\alpha)}_0-E^{(\beta)}_0}\ket{\beta},
\end{equation}
where $\ket{\alpha}$ and $\ket{\beta}$ are eigenstates of the Fibonacci Hamiltonian $H_0$, and the $E_0$'s are eigenvalues. We expand these states in the position basis, $\ket{\alpha}=\sum_i\alpha(x_i)\ket{i}$, and obtain (omitting normalization for now)
\begin{align*}
    \ket{\psi}&=\sum_{i=1}^N\alpha(x_i)\ket{i}+\sum_{\beta\neq\alpha}\sum_{i,j,k}\frac{\beta^*(x_i)\alpha(x_j)\beta(x_k)\bra{i}V\ket{j}}{E^{(\alpha)}_0-E^{(\beta)}_0}\ket{k} \\
    &=\sum_{i=1}^N\alpha(x_i)\ket{i}+V_I\sum_{\beta\neq\alpha}\sum_{k=1}^N\frac{\beta^*(x_m)\alpha(x_m)\beta(x_k)}{E^{(\alpha)}_0-E^{(\beta)}_0}\ket{k} \\
    &=\sum_{i=1}^N\left(\alpha(x_i)+V_I\sum_{\beta\neq\alpha}\frac{\beta^*(x_m)\alpha(x_m)\beta(x_i)}{E^{(\alpha)}_0-E^{(\beta)}_0}\right)\ket{i} \\
    &\equiv \sum_{i=1}^N\psi^m_\alpha(x_i)\ket{i}.
\end{align*}
The second line results from the matrix elements of the perturbation $V$ only having one element $V_I$ on the diagonal, corresponding to site $m$. In the last line, we defined our quantity of interest $\psi^m_\alpha(x_i)$, the amplitude of the states at sites $x_i$. The subscript $\alpha$ is a reminder that its zeroth order state is $\ket{\alpha}$ and the superscript $m$ indicates that the impurity has been placed at the site $x_m$. We also need to know the norm of the state to properly normalize it. At this order of perturbation theory, it is given by
\begin{align*}
    \bra{\psi}\ket{\psi}&=1+\sum_{\substack{\beta\neq\alpha \\ \beta'\neq\alpha}}\frac{\bra{\alpha}V\ket{\beta}\bra{\beta'}V\ket{\alpha}}{(E^{\alpha}_0-E^{\beta}_0)(E^{\alpha}_0-E^{\beta'}_0)}\bra{\beta}\ket{\beta'} \\
    &=1+\sum_{\beta\neq\alpha}\frac{|\bra{\alpha}V\ket{\beta}|^2}{(E^{(\alpha)}_0-E^{(\beta)}_0)^2}\\
    &=1+V_I\sum_{\beta\neq\alpha}\frac{\big|\alpha^*(x_m)\beta(x_m)\big|^2}{(E^{(\alpha)}_0-E^{(\beta)}_0)^2}.
\end{align*}
With the state and its norm, we can write down an approximate expression for the IPR
\begin{equation}
    I_\alpha=\frac{\sum_{i=1}^N|\psi^m_\alpha(x_i)|^2}{\sum_{i=1}^N|\psi^m_\alpha(x_i)|^4},
\end{equation}
where we assume proper normalization of the state $\ket{\psi}$. We start by working out the expression for $|\psi_\alpha^m(x_i)|^4$ to $2^\text{nd}$ order in $V_I$,
\begin{widetext}
    \begin{align}
        |\psi_\alpha^m(x_i)|^4&=\left[\left(\alpha(x_i)+V_I\sum_{\beta\neq\alpha}\frac{\beta^*(x_m)\alpha(x_m)\beta(x_i)}{E^{(\alpha)}_0-E^{(\beta)}_0}\right)\left(\alpha^*(x_i)+V_I\sum_{\beta\neq\alpha}\frac{\beta(x_m)\alpha^*(x_m)\beta^*(x_i)}{E^{(\alpha)}_0-E^{(\beta)}_0}\right)\right]^2 \notag \\
        &=|\alpha(x_i)|^4+4|\alpha(x_i)|^2\text{Re}\big[f^m_\alpha(x_i)\big]V_I+\bigg[4\big|f^m_\alpha(x_i)\big|^2+2\text{Re}\big[(f^m_\alpha(x_i))^2\big]\bigg]V_I^2 +\mathcal{O}(V_I^3). 
    \end{align}
\end{widetext}
where we defined the function 
\begin{equation}
    f^m_\alpha(x_i)\equiv\alpha^*(x_i)\sum_{\beta\neq\alpha}\frac{\alpha(x_m)\beta^*(x_m)\beta(x_i)}{E^{(\alpha)}_0-E^{(\beta)}_0}.
\end{equation}
This function is the quantity that explains why the grouping in terms of the renormalization path of the sites happens. As mentioned previously in the article, the amplitudes $\alpha_(x_m)$ and $\beta(x_m)$ have most of their support on the sites belonging to the same renormalization path.
% For $|\psi^m_\alpha(x_i)|^2$, we have 
% \begin{equation}
%     |\psi^m_\alpha(x_i)|^2=|\alpha(x_i)|^2+\text{Re}\big[f^m_\alpha(x_i)\big]V_I+\frac{|f^m_\alpha(x_i)|^2}{|\alpha(x_m)|^2}V_I^2.
% \end{equation}
With this, we have an analytic expression for the IPR of state $\ket{\psi_\alpha}$,
\begin{equation}\label{IPRbef}
    I_\alpha=\frac{1+\sum_i\left(\text{Re}\big[f^m_\alpha(x_i)\big]V_I+\frac{|f^m_\alpha(x_i)|f^m_\alpha(x_i)|^2}{|\alpha(x_i)|^2}V_I^2\right)}{\mathcal{I}^{-1}_\alpha+\zeta^m_\alpha V_I+\gamma^m_\alpha V_I^2},
\end{equation}
where we defined three other expressions for the coefficients in front of the impurity strengths for the $|\psi_\alpha^m(x_i)|^4$ expression:
\begin{align*}
    \mathcal{I}^{-1}_\alpha&\equiv\sum_i|\alpha(x_i)|^4, \\
    \zeta^m_\alpha&\equiv\sum_i4|\alpha(x_i)|^2\text{Re}\big[f^m_\alpha(x_i)\big], \\
    \gamma^m_\alpha&\equiv\sum_i\bigg[4\big|f^m_\alpha(x_i)\big|^2+2\text{Re}\big[(f^m_\alpha(x_i))^2\big]\bigg].
\end{align*}
% \begin{align}
%     \mathcal{I}^{-1}_\alpha&=\sum_i|\alpha(x_i)|^4 \\
%     \zeta^m_\alpha&\equiv\sum_i4|\alpha(x_i)|^2\text{Re}\big[f^m_\alpha(x_i)\big], \\
%     \gamma^m_\alpha&\equiv\sum_i\bigg[4\big|f^m_\alpha(x_i)\big|^2+2\text{Re}\big[(f^m_\alpha(x_i))^2\big]\bigg].
% \end{align}
% We should note that in Eq.\eqref{IPRbef}, the numerator is exact in terms of $V_I$, while the denominator has been written up to $V_I^2$ order. We further expand the expression for the IPR to have it as a second degree polynomial in $V_I$, yielding 
% \begin{widetext}
% \begin{equation}
%     I_\alpha=\mathcal{I}_\alpha\left\{1+\left[\sum_i\text{Re}\big[f^m_\alpha(x_i)\big]-\mathcal{I}_\alpha\zeta^m_\alpha\right]V_I+\left[\sum_i\left(\frac{|f^m_\alpha(x_i)|^2}{|\alpha(x_i)|^2}-\mathcal{I}_\alpha\zeta^m_\alpha\text{Re}\big[f^m_\alpha(x_i)\big]\right)+\frac{(\zeta^m_\alpha)^2-\gamma^m_\alpha\mathcal{I}_\alpha^{-1}}{\mathcal{I}_\alpha^2}\right]V_I^2\right\}.
% \end{equation}
% \end{widetext}

\end{document}